\documentclass[trackchanges, twocolumn, resetfootnote]{aastex701}
\usepackage{CJK}
\newcommand{\ngc}{NGC\,7213}

\newcommand{\ms}{\ensuremath{M_{\odot}}}

\newcommand{\kms}{\ensuremath{\mathrm{km}\,\mathrm{s}^{-1}}}

\newcommand{\rg}{\ensuremath{R_{\rm g}}}

\begin{document}

\title{XRISM/Resolve reveals the complex iron structure of NGC\,7213:\\ Evidence for radial stratification between inner disk and broad-line region}

\author[orcid=0000-0002-0273-218X]{E. Kammoun}
\affiliation{Cahill Center for Astronomy \& Astrophysics, California Institute of Technology, 1216 East California Boulevard, Pasadena, CA 91125, USA}
\email[show]{ekammoun@caltech.edu} 

\author[orcid=0000-0002-6808-2052]{T. Kawamuro} 
\affiliation{Department of Earth and Space Science, Osaka University, 1-1 Machikaneyama, Toyonaka 560-0043, Osaka, Japan}
\email{kawamuro@ess.sci.osaka-u.ac.jp}

\author{K. Murakami} 
\affiliation{Department of Earth and Space Science, Osaka University, 1-1 Machikaneyama, Toyonaka 560-0043, Osaka, Japan}
\email{murakami@ess.sci.osaka-u.ac.jp}

\author[orcid=0000-0002-4622-4240]{S. Bianchi} 
\affiliation{Dipartimento di Matematica e Fisica, Universita degli Studi Roma Tre, via della Vasca Navale 84, I-00146 Roma, Italy}
\email{stefano.bianchi@uniroma3.it}

\author[orcid=0000-0002-6896-1364]{F. Nicastro} 
\affiliation{INAF - Osservatorio Astronomico di Roma Via Frascati, 33 I-00078 Monte Porzio Catone (RM), Italy}
\email{fabrizio.nicastro@inaf.it}

\author[orcid=0000-0002-1035-8618]{A. Luminari} 
\affiliation{INAF – Istituto di Astrofisica e Planetologia Spaziali, Via del Fosso del Caveliere 100, I-00133 Roma, Italy}
\affiliation{INAF - Osservatorio Astronomico di Roma Via Frascati, 33 I-00078 Monte Porzio Catone (RM), Italy}
\email{alfredo.luminari@inaf.it}

\author[orcid=0000-0001-8525-3442]{E. Aydi} 
\affiliation{Department of Physics \& Astronomy, Texas Tech University, Box 41051, Lubbock, TX 79409-1051, USA}
\email{eaydi@ttu.edu}

\author[orcid=0000-0002-3719-940X]{M. Eracleous}
\affiliation{Department of Astronomy \& Astrophysics and Institute for Gravitation and the Cosmos, 525 Davey Lab, 251 Pollock Road, The Pennsylvania State University, University Park, PA 16802, USA}
\email{mxe17@psu.edu}


\author[orcid=0000-0002-5966-4210]{O. K. Adegoke}
\affiliation{Cahill Center for Astronomy \& Astrophysics, California Institute of Technology, Pasadena, 1216 East California Boulevard, CA 91125, USA}
\email{oadegoke@caltech.edu}

\author[orcid=0000-0001-5487-2830]{E. Bertola} 
\affiliation{INAF - Osservatorio Astrofisco di Arcetri, largo E. Fermi 5, 50127, Firenze, Italy}
\email{elena.bertola@inaf.it}

\author[orcid=0000-0001-9379-4716]{P. G. Boorman}
\affiliation{Cahill Center for Astronomy \& Astrophysics, California Institute of Technology, Pasadena, 1216 East California Boulevard, CA 91125, USA}
\email{boorman@caltech.edu}

\author[orcid=0000-0002-2629-4989]{V. Braito}
\affiliation{INAF - Osservatorio Astronomico di Brera, Via Bianchi 46 I-23807 Merate (LC), Italy}
\affiliation{Department of Physics, Institute for Astrophysics and Computational Sciences, The Catholic University of America, Washington, DC 20064, USA}
\affiliation{Dipartimento di Fisica, Universit\'a di Trento, Via Sommarive 14, 38123 Trento, Italy}
\email{valentina.braito@inaf.it}

\author[orcid=0000-0002-5182-6289]{G. Bruni} 
\affiliation{INAF – Istituto di Astrofisica e Planetologia Spaziali, Via del Fosso del Caveliere 100, I-00133 Roma, Italy}
\email{gabriele.bruni@inaf.it}

\author[orcid=0000-0003-3451-9970]{A. Comastri}
\affiliation{INAF - Osservatorio di Astrofisica e Scienza dello Spazio (OAS) di Bologna, via P. Gobetti 93/3, I-40129 Bologna, Italy}
\email{andrea.comastri@inaf.it}

\author{P. Cond\`{o}}
\affiliation{Dipartimento di Fisica, Università degli Studi di Roma ``Tor Vergata'', via della Ricerca Scientifica 1, I-00133 Roma, Italy}
\email{pierpaolo.condo@roma2.infn.it}

\author{M. Dadina}
\affiliation{INAF - Osservatorio di Astrofisica e Scienza dello Spazio (OAS) di Bologna, via P. Gobetti 93/3, I-40129 Bologna, Italy}
\email{mauro.dadina@inaf.it}

\author[]{T. Enoto}
\affiliation{Department of Physics, Kyoto University, Kitashirakawa Oiwake, Sakyo, Kyoto 606-8502, Japan}
\email{enoto.teruaki.2w@kyoto-u.ac.jp}

\author[0000-0003-3828-2448]{J. A. Garc\'ia}
\affiliation{X-ray Astrophysics Laboratory, NASA Goddard Space Flight Center, Greenbelt, MD 20771, USA}
\affiliation{Cahill Center for Astronomy \& Astrophysics, California Institute of Technology,
Pasadena, CA 91125, USA}
\email{javier@caltech.edu}

\author[orcid=0000-0002-9719-8740]{V. E. Gianolli} 
\affiliation{Department of Physics and Astronomy, Clemson University, Kinard Lab of Physics, Clemson, SC 29634, USA}
\email{vgianol@clemson.edu}

\author[orcid=0000-0002-4226-8959]{F. A. Harrison}
\affiliation{Cahill Center for Astronomy \& Astrophysics, California Institute of Technology, Pasadena, 1216 East California Boulevard, CA 91125, USA}
\email{fiona@srl.caltech.edu}

\author[orcid=0000-0001-9094-0984]{G. Lanzuisi}
\affiliation{INAF - Osservatorio di Astrofisica e Scienza dello Spazio (OAS) di Bologna, via P. Gobetti 93/3, I-40129 Bologna, Italy}
\email{giorgio.lanzuisi@inaf.it}

\author[orcid=0000-0001-5762-6360]{M. Laurenti} 
\affiliation{Dipartimento di Fisica, Università degli Studi di Roma ``Tor Vergata'', via della Ricerca Scientifica 1, I-00133 Roma, Italy}
\affiliation{INFN – Sezione di Roma ``Tor Vergata'', Via della Ricerca Scientifica 1, I-00133 Roma, Italy}
\affiliation{INAF - Osservatorio Astronomico di Roma Via Frascati, 33 I-00078 Monte Porzio Catone (RM), Italy}
\email{marco.laurenti@roma2.infn.it}

\author[orcid=0000-0002-2055-4946]{A. Marinucci}
\affiliation{ASI – Agenzia Spaziale Italiana, Via del Politecnico snc, 00133 Roma, Italy}
\email{andrea.marinucci@asi.it}

\author[0000-0003-4216-7936]{G. Mastroserio}
\affiliation{Scuola Universitaria Superiore IUSS Pavia, Palazzo del Broletto, piazza della Vittoria 15, I-27100 Pavia, Italy}
\email{guglielmo.mastroserio@iusspavia.it}

\author[]{H. Matsumoto}
\affiliation{Department of Earth and Space Science, Osaka University, 1-1 Machikaneyama, Toyonaka 560-0043, Osaka, Japan}
\email{matumoto@ess.sci.osaka-u.ac.jp}

\author[orcid=0000-0002-2152-0916]{G. Matt}
\affiliation{Dipartimento di Matematica e Fisica, Universita degli Studi Roma Tre, via della Vasca Navale 84, I-00146 Roma, Italy}
\email{giorgio.matt@uniroma3.it}

\author[orcid=0000-0003-1994-5322]{G. Matzeu}
\affiliation{European Space Agency (ESA), European Space Astronomy Centre
(ESAC), 28691 Villanueva de la Cañada, Madrid, Spain}
\email{Gabriele.Matzeu@ext.esa.int}

\author[orcid=0000-0001-9815-9092]{R. Middei}
\affiliation{INAF - Osservatorio Astronomico di Roma Via Frascati, 33 I-00078 Monte Porzio Catone (RM), Italy}
\affiliation{Space Science Data Center, Agenzia Spaziale Italiana, Via del Politecnico snc, 00133 Roma, Italy}
\email{riccardo.middei@ssdc.asi.it}

\author[orcid=0000-0001-9226-8992]{E. Nardini}
\affiliation{INAF - Osservatorio Astrofisco di Arcetri, largo E. Fermi 5, 50127, Firenze, Italy}
\email{emanuele.nardini@inaf.it}

\author[orcid=0000-0001-6020-517X]{H. Noda}
\affiliation{Astronomical Institute, Tohoku University, Miyagi 980-8578, Japan}
\email{hirofumi.noda@astr.tohoku.ac.jp}

\author[]{H. Odaka}
\affiliation{Department of Earth and Space Science, Osaka University, 1-1 Machikaneyama, Toyonaka 560-0043, Osaka, Japan}
\email{odaka@ess.sci.osaka-u.ac.jp}

\author[]{S. Ogawa}
\affiliation{Institute of Space and Astronautical Science (ISAS), Japan Aerospace Exploration Agency (JAXA), Kanagawa 252-5210, Japan}
\email{sogawa@ac.jaxa.jp}

\author[orcid=0000-0003-0543-3617]{F. Panessa} 
\affiliation{INAF – Istituto di Astrofisica e Planetologia Spaziali, Via del Fosso del Caveliere 100, I-00133 Roma, Italy}
\email{francesca.panessa@inaf.it}

\author[orcid=0000-0001-9095-2782]{E. Piconcelli}
\affiliation{INAF - Osservatorio Astronomico di Roma Via Frascati, 33 I-00078 Monte Porzio Catone (RM), Italy}
\email{enrico.piconcelli@inaf.it}

\author[orcid=0000-0003-2532-7379]{C. Pinto}
\affiliation{INAF/IASF Palermo, via Ugo La Malfa 153, I-90146 Palermo, Italy}
\email{ciro.pinto@inaf.it}

\author[orcid=0000-0003-1661-2338]{J. M. Piotrowska}
\affiliation{Cahill Center for Astronomy \& Astrophysics, California Institute of Technology, Pasadena, 1216 East California Boulevard, CA 91125, USA}
\email{joannapk@caltech.edu}

\author[orcid=0000-0003-0293-3608]{G. Ponti} 
\affiliation{INAF - Osservatorio Astronomico di Brera, Via Bianchi 46 I-23807 Merate (LC), Italy}
\affiliation{Max-Planck-Institut f\"{u}r extraterrestrische Physik, Gie{\ss}enbachstra{\ss}e 1, 85748, Garching, Germany}
\affiliation{Como Lake Center for Astrophysics (CLAP), DiSAT, Università degli Studi dell'Insubria, via Valleggio 11, 22100 Como, Italy}
\email{gabriele.ponti@inaf.it}

\author[orcid=0000-0001-5231-2645]{C. Ricci}
\affiliation{Department of Astronomy, University of Geneva, ch. d’Ecogia 16, 1290, Versoix, Switzerland}
\email{claudio.ricci.astro@gmail.com}

\author[orcid=0000-0003-4631-1528]{R. Ricci}
\affiliation{Dipartimento di Fisica, Università degli Studi di Roma ``Tor Vergata'', via della Ricerca Scientifica 1, I-00133 Roma, Italy}
\affiliation{INAF-Istituto di Radioastronomia, Via Gobetti 101, I-40129 Bologna, Italy}
\email{roberto.ricci@inaf.it}

\author[orcid=0000-0003-1200-5071]{R. Serafinelli} 
\affiliation{Instituto de Estudios Astrof\'{i}sicos, Facultad de Ingenier\'{i}a y Ciencias, Universidad Diego Portales, Avenida Ej\'{e}rcito Libertador 441, Santiago, Chile}
\affiliation{INAF - Osservatorio Astronomico di Roma Via Frascati, 33 I-00078 Monte Porzio Catone (RM), Italy}
\email{roberto.serafinelli@mail.udp.cl}

\author[orcid=0000-0003-3922-5007]{F. Shi}
\affiliation{Shanghai Astronomical Observatory, Chinese Academy of Science, No.80 Nandan Road, Shanghai, China}
\email{fzshi@shao.ac.cn}

\author[orcid=0000-0003-2686-9241]{D. Stern}
\affiliation{Jet Propulsion Laboratory, California Institute of Technology, Pasadena, CA 91109, USA}
\email{daniel.k.stern@jpl.nasa.gov}

\author[]{A. Tanimoto}
\affiliation{Graduate School of Science and Engineering, Kagoshima University, Kagoshima, 890-8580, Japan}
\email{ats141592@gmail.com}

\author[orcid=0000-0003-1780-5481]{Y. Terashima}
\affiliation{Department of Physics, Ehime University, Ehime 790-8577, Japan}
\email{terashima.yuichi.mc@ehime-u.ac.jp}

\author[]{R. Tomaru}
\affiliation{Department of Earth and Space Science, Osaka University, 1-1 Machikaneyama, Toyonaka 560-0043, Osaka, Japan}
\email{tomaru@ess.sci.osaka-u.ac.jp}
 
\author[orcid=0000-0002-6562-8654]{F. Tombesi} 
\affiliation{Dipartimento di Fisica, Università degli Studi di Roma ``Tor Vergata'', via della Ricerca Scientifica 1, I-00133 Roma, Italy}
\affiliation{INAF - Osservatorio Astronomico di Roma Via Frascati, 33 I-00078 Monte Porzio Catone (RM), Italy}
\affiliation{INFN – Sezione di Roma ``Tor Vergata'', Via della Ricerca Scientifica 1, I-00133 Roma, Italy}
\email{francesco.tombesi@roma2.infn.it}

\author[orcid=0000-0003-3450-6483]{A. Tortosa}
\affiliation{INAF - Osservatorio Astronomico di Roma Via Frascati, 33 I-00078 Monte Porzio Catone (RM), Italy}
\email{alessia.tortosa@inaf.it}

\author[]{Y. Ueda}
\affiliation{Department of Astronomy, Kyoto University, Kyoto 606-8502, Japan}
\email{ueda@kusastro.kyoto-u.ac.jp}

\author[orcid=0000-0001-9442-7897]{F. Ursini} 
\affiliation{Dipartimento di Matematica e Fisica, Universita degli Studi Roma Tre, via della Vasca Navale 84, I-00146 Roma, Italy}
\email{francesco.ursini2@uniroma3.it}

\author[orcid=0000-0002-8853-9611]{C. Vignali}
\affiliation{Dipartimento di Fisica e Astronomia ``Augusto Righi'', Università degli Studi di Bologna, via P. Gobetti 93/2, 40129 Bologna, Italy}
\affiliation{INAF - Osservatorio di Astrofisica e Scienza dello Spazio (OAS) di Bologna, via P. Gobetti 93/3, I-40129 Bologna, Italy}
\email{cristian.vignali@unibo.it}

\author[orcid=0000-0002-9754-3081]{S. Yamada}
\affiliation{Frontier Research Institute for Interdisciplinary Sciences, Tohoku University, Sendai 980-8578, Japan}
\email{satoshi.yamada@riken.jp}

\author[orcid=0000-0003-4808-893X]{S. Yamada}
\affiliation{Department of Physics, Rikkyo University, 3-34-1 Nishi Ikebukuro, Toshima-ku, Tokyo 171-8501, Japan}
\email{syamada@rikkyo.ac.jp}


\begin{abstract}

We present the first high-resolution X-ray spectrum of \ngc\ obtained with XRISM/Resolve, supported by simultaneous XMM-Newton, NuSTAR, and SOAR optical data. The XRISM spectrum resolves the neutral Fe\,K$\alpha$ into two components: a narrow core ($\rm FWHM = 650_{-220}^{+240}\,\kms$) consistent with emission at the dust sublimation radius, and a broader, asymmetric line best described by disk-like emission from $\sim100\,\rg$. This disk component mirrors the profile of the double-peaked H$\alpha$ line observed in the optical. In addition, we detect broadened \ion{Fe}{25} and \ion{Fe}{26} emission lines whose inferred locations bridge the gap between the inner disk and the optical broad-line region. The weak narrow Fe K$\alpha$ equivalent width ($\rm EW = 32 \pm 6\,eV$) and absence of a Compton hump imply a low-covering-fraction, Compton-thin torus. Together, these results reveal a radially stratified structure in NGC\,7213, spanning nearly four orders of magnitude in radius, and place the source in an intermediate accretion state ($\lambda_ {\rm Edd} = 0.1–1\,\%$) where the inner disk and BLR remain, while the torus shows signs of dissipation.

\end{abstract}


\section{Introduction}

One of the main observational signatures of active galactic nuclei (AGN) is the broadening of atomic emission lines by high-velocity motion of gas near the central supermassive black hole (SMBH). Over decades, extensive efforts have been dedicated to characterizing the broad-line region (BLR) emission lines, kinematics, and ionization structure, informing the interplay between the SMBH and its surrounding gas. BLR observations can also constrain the mechanism for transporting gas inwards through accretion disks or outwards through winds \citep{Krolik2001}. The source of the broad emission lines has been associated in a variety of works with the surface of the accretion disk or a wind arising at the surface of the outer accretion disk \citep[e.g.,][and references therein]{Collin-Souffrin1987,Collin-Souffrin1988, Emmering1992, Murray1997, Baskin2018,Naddaf2021}.  A characteristic length scale for the BLR is commonly estimated by measuring the delay between the variations of brightness of the accretion disk continuum and the corresponding variations of the broad emission lines \citep[known as reverberation mapping;][]{Blandford1982}. Directly imaging such regions has been challenging because of their small angular size ($\lesssim 0.1\,\rm mas$). It is only very recently that the BLR could be spatially resolved in a handful of bright, nearby AGN thanks to the development of a new generation of optical/near-infrared (O/NIR) interferometers \citep{GRAVITY2024}. However, despite the wealth of knowledge gained from UV/O/NIR observations, a comprehensive understanding of the BLR's complex dynamics and ionization processes remains elusive. 

Studying the BLR has predominantly been conducted using UV/optical spectroscopy. In X-rays, the Fe\,K$\alpha$ line could be spectroscopically resolved in a limited number of bright AGN using the Chandra High-Energy Transmission Grating (HETG). Chandra/HETG observations suggest that there is no universal location of the Fe\,K$\alpha$ line-emitting region relative to the optical BLR. In general, a given source may have contributions to the Fe\,K$\alpha$ line originating at parsec-scale distances from the SMBH, down to distances a factor $\sim 2$ closer to the SMBH than the BLR H$\beta$ \citep{Yaqoob2001, Shu2010}. Only in a few cases has it been possible to show that the Fe\,K$\alpha$ line width measured by Chandra/HETG is produced in the optical BLR \citep[][]{Bianchi2008, Shu2010, Miller2018}. Recent studies have shown that, for local AGN, the Fe\,K$\alpha$ emitting region size is smaller than the dust sublimation radius in the  majority of the sources \citep[e.g.][]{Gandhi2015, Andonie2022}. The Resolve X-ray spectrometer onboard the X-ray Imaging and Spectroscopy Mission \citep[XRISM][]{Tashiro2024} is providing an unprecedented look at the BLR in X-rays \citep[e.g.,][]{Xrism2024ngc4151, Bogensberger2025, Miller2025_mrk279}.

Over the last three decades, an increasing number of AGN has been found to exhibit long-lived broad ($\sim 10,000 \rm\, km\,s^{-1}$) double-peaked Balmer emission lines \citep[e.g.,][]{Eracleous1994, Eracleous2003, Strateva2003, Storchi-Bergmann2017, Ward2025}. These lines are commonly modeled as emission from a relativistic, Keplerian disk-like structure, where Doppler boosting results in asymmetry between the red and blue peaks \citep[see e.g.,][]{Chen1989a, Chen1989b, Storchi-Bergmann1993, Eracleous1994, Eracleous2003, Strateva2003, Bianchi2019, Bianchi2022, Ward2024, Ward2025}. \cite{Eracleous1994} estimated that double-peaked emitters make up $\sim 15\%$ of radio-loud AGN at $z < 0.4$. \cite{Strateva2003}  estimated that double-peaked
emitters make up  $\sim 4\%$ of the SDSS (mostly radio-quiet) quasar population at $z < 0.4$, while \cite{Ward2025} estimated that double-peaked emitters make up $\sim 21\%$ of all X-ray selected
low-redshift AGN at $z < 0.3$ in the BASS sample. Some of these sources show substantial changes in the relative flux of the blue and red peaks over timescales of months to years, which is well modeled by the rotation of spiral arms or hot spots in the disk \citep[e.g.,][]{Gezari2007, Lewis2010, Ward2024}. If these double-peaked features originate from the accretion disk, we should expect to see analogous Fe lines in the X-ray spectra of these sources.

\subsection{NGC 7213}
Located at a distance of 22.8\,Mpc, the nearly face-on spiral galaxy \ngc\ harbors an intermediate/low-luminosity AGN powered by a BH with a mass of $8_{-6}^{+16} \times 10^7\,M_\odot$ \citep[derived using velocity dispersion;][]{Schnorr-Muller2014}, accreting at $\sim 0.05-1\%$ of its Eddington limit. \ngc\ has been proposed as a high-mass analog of hard-state X-ray binaries \citep[XRBs;][]{Emmanoulopoulos2012}, showing a `harder when brighter' behavior. The source lies between the radio-loud and radio-quiet regimes, showing no extended radio emission. \cite{Bell2011} showed that the X-ray and radio light curves of the source are correlated, with the radio variability lagging behind the X-rays by 24\,d (48\,d) at 8.4\,GHz (4.8\,GHz). The available data suggest that the source could either be analogous to radio-loud AGN where the jet is powered by an advection-dominated accretion flow, or it is similar to a hard-state XRB with a truncated accretion disk. 

\cite{Bianchi2008} described the Chandra/HETG spectrum with a power law ($\Gamma = 1.69 \pm 0.01$) and three emission lines. The strongest of these lines is consistent with a neutral Fe\,K$\alpha$ at $6.397^{+0.006}_{-0.011}\,\rm keV$ with a full width at half maximum $\rm FWHM = 2400_{-600}^{+1100}\,\kms$ ($\sigma = 22^{+10}_{-6}\,\rm eV$), and an equivalent width of $\rm EW = 120^{+40}_{-30}\,eV$. The broadening of this line is consistent with the value measured for the broad H$\alpha$ emission line of $2640^{+110}_{-90}\,\kms$. In addition, two ionized Fe lines consistent with the He-like \ion{Fe}{25} and H-like \ion{Fe}{26} are hinted at in the Chandra/HETG spectrum. \cite{Bianchi2008} reported a tentative blueshift of the \ion{Fe}{25} and \ion{Fe}{26} lines by around $1000\,\kms$, which they interpret as a possible indication of starburst winds. \cite{Shi2022} attributed this tentative blueshift to hot winds from the accretion flow \citep[see also][]{Shi2024}.

XMM-Newton and Suzaku observations of the source confirmed the Chandra/HETG results, albeit at lower resolution \citep{Lobban2010}. The broadening and the EW of the low-ionization Fe\,K line both suggest that this line originates from Compton-thin gas at a distance consistent with the BLR, although with large uncertainties. Interestingly, the NuSTAR observation of \ngc\ showed that its hard X-ray spectrum can be described by a power law with a high-energy cutoff, showing no signature of a Compton hump \citep{Ursini2015}. This further supports the conclusion that the low-ionization Fe\,K line is produced by Compton-thin material with an equivalent hydrogen column density $N_{\rm H} = 5.0^{+2.0}_{-1.6}\times 10^{23}\,\rm cm^{-2}$.

The optical spectrum of \ngc\ revealed the presence of a double-peaked broad (DPB) H$\alpha$ emission line in addition to the regular symmetric BLR-like line \citep{Schnorr-Muller2014, Schimoia2017}. \cite{Schimoia2017} studied the variability of the H$\alpha$ complex, and showed that both parts of the line profile are variable. The authors estimated the evolution timescales of the symmetric part to be of $\sim 7-21$\,days, and that of the DPB part to be of the order of a few months. This suggests that the symmetric part evolves on a light-travel time scale (consistent with the distance between the BLR and the central engine), while the DPB part is more dominated by slower dynamical time scales. They modeled the DPB part assuming that it originates from a region of a Keplerian and relativistic accretion disk. Their model resulted in an DPB emitting region between $\sim 300\,\rg$ and $3000\,\rg$, where $\rg = GM_{\rm BH}/c^2$ is the gravitational radius, with an inclination of $47\degr \pm 2\degr$. Within the context of their model, the disk harbors a spiral arm with varying contrast relative to the underlying disk, which processes causes the variability in the DPB part of the profile.

In this paper, we present the first high spectral resolution observation of \ngc\ with XRISM. Simultaneous observations with XMM-Newton, NuSTAR, and other ground-based facilities have been performed to support the XRISM observation. In this first paper we focus on the Fe lines region ($\sim 6-7$\,keV range), employing mainly phenomenological models. Follow-up papers will discuss in greater detail the broad-band aspect of these spectra as well as testing a variety of physically motivated models.

In Section\,\ref{sec:obs}, we present the details of the observation and the data reduction procedures. The results of the spectral analysis are described in Section\,\ref{section:spectra}. We discuss our findings and their implications in Section\,\ref{sec:discussion}. We present our conclusions in Section\,\ref{sec:conclusion}.

\section{Observations and data reduction}
\label{sec:obs}
\subsection{XRISM observation}

XRISM observed \ngc\ on 2024 November 4-7 for a total of 118.5\,ks (ObsID 201115010; PIs: E. Kammoun and T. Kawamuro). XRISM is a JAXA/NASA collaborative mission, with ESA participation, consisting of two instruments: a high-resolution spectrometer (Resolve; \citealt{Resolve2025}) and a soft X-ray imaging spectrometer (Xtend; \citealt{Xtend2025}). The Resolve and Xtend instruments were operated in ``PX\_NORMAL'' mode and ``1/8 window'' mode, respectively. In this work, we focus only on the Resolve data. We defer the full analysis of the Xtend data to a subsequent paper. 

We reprocessed the Resolve data using the \textsc{xapipeline} command and the latest calibration files (\textsc{CalDB 11}, released on 2025-03-11). We produced the level-2 cleaned files following the XRISM data reduction ABC guide\footnote{\url{https://heasarc.gsfc.nasa.gov/docs/xrism/analysis/abc_guide/xrism_abc.html}} v1.0. We applied the recommended screening for pulse rise time, event type and status. In addition, we excluded the events from pixel number 27, as recommended to avoid calibration uncertainties. We also applied a filter of the geomagnetic cut-off rigidity (COR), which is roughly inversely proportional to the particle background in low Earth orbits. We tested different COR thresholds and we confirmed that the results did not significantly change as discussed in Appendix\,\ref{appendix:COR}. We used a threshold of $\rm COR > 6$, which reduced the net exposure time to 90.3\,ks. Our choice can be considered as a reasonable compromise between exposure time and S/N. We selected only the high-resolution events (Hp) which accounted for 95.2\% of all the events during this observation.

We used the \textsc{XSELECT} task to extract spectra and light curves. Then, we used the \textsc{rslmkrmf} command to generate the extra-large (XL) type of response file for Resolve. Next, we created a standard exposure map for both instruments using the \textsc{xaexpmap} command. We finally created the ancillary response file (ARF) using the \textsc{xaarfgen} command. This command runs a ray-tracing simulation to calculate the reflection and transmission of photons from the assumed X-ray source, and to count the photons detected in the selected pixels. We set the number of photons in the simulations \textsc{numphoton=600,000} and limited the ARF to the 2-10\,keV range. Given the brightness of the source and the low non-X-ray background (NXB) of Resolve, we can safely neglect the NXB during the spectral fits. As a check, we verified that no signature of the known NXB lines in the spectrum of the source.

\subsection{XMM-Newton observations}

Two XMM-Newton Director's Discretionary Time (DDT) observations of \ngc\ took place on 2024 November 5 and on 2024 November 6 (ObsIDs  0953012201 and  0953012301, respectively; PI: E. Kammoun), for total exposure times of 47.4\,ks and 56.4\,ks. We refer to these observations as XMM-Newton ObsA and ObsB, respectively, hereafter. A problem in the communication with the ground station occurred during the first observation, which resulted in a shorter exposure. The observations were operated in the Large Window/Thin Filter mode for EPIC-pn \citep{Stru01} and Small Window/Thin Filter mode for the two EPIC-MOS \citep{Tur01} instruments. In this work, we focus on the EPIC-pn data only\footnote{The EPIC-MOS and EPIC-pn spectra are consistent. For the purposes of this work, including EPIC-MOS spectra does not affect any of the conclusions.}. We use the XMM-Newton Science Analysis System (\textsc{SAS v22.1.0}) to reduce and analyze the data of these observations. Source spectra and light curves were extracted from an annular region with an inner radius of 10\arcsec\ and an outer radius of 35\arcsec\ centered on the source to reduce pile-up. The corresponding background spectra were extracted from an off-source circular region located on the same CCD chip, with a radius of 75\arcsec. We filtered out periods with strong background flares. This resulted in net exposures of 9.9\,ks and 29.6\,ks, respectively. Response matrices were produced using the \textsc{FTOOLs} \textsc{RMFGEN} and \textsc{ARFGEN}. The XMM-Newton light curves do not show any significant variability.

\subsection{NuSTAR observation}

A NuSTAR DDT observation of \ngc\ took place on 2024 November 5 (ObsID 91001636002; PI: E. Kammoun) for a total exposure of 52\,ks. The data were reduced using the standard pipeline in the NuSTAR Data Analysis Software (\textsc{NuSTARDAS v2.1.4}), using \textsc{CalDB 20241104}. We cleaned the unfiltered event files with the standard depth correction. We reprocessed the data using the \textsc{saamode = optimized}, \textsc{saacalc=3}, and \textsc{tentacle = yes} criteria for a more conservative treatment of the high background levels near the South Atlantic Anomaly. We used the \textsc{optimal-radius} tool in \textsc{nustar-gen-utils} \citep{nustar-gen-utils} to estimate the optimal source region to maximize the signal-to-noise ratio above 10\,keV. Thus, we extracted the source and background light curves and spectra from circular regions of radii 51\arcsec\ and 90\arcsec, respectively, for the two focal plane modules (FPMA and FPMB) using the HEASOFT task \textsc{nuproducts}. In the following, we analyze the spectra from FPMA and FPMB jointly, without combining them. All X-ray spectra in this work were binned using the ftool \textsc{ftgrouppha} with optimal binning \citep{kaastra16}.

\subsection{SOAR observation}

On 2024 November 2, we obtained medium-resolution optical spectroscopy for \ngc\ using the Goodman spectrograph \citep{Clemens2004} on the 4.1\,m Southern Astrophysical Research (SOAR) telescope located on Cerro Pach\'on, Chile. We used a setup with a 2100~l\,mm$^{-1}$ grating and a 0.95\arcsec\ slit, yielding a resolution $R \approx 5000$ in a region centered on H$\alpha$ and H$\beta$ that is 570\,\AA\ wide. The spectra were reduced and optimally extracted using the \textsc{apall} package in the Image Reduction and Analysis Facility \citep[IRAF;][]{Tody1986}. In this work, we focus on the H$\alpha$ region. The results from the H$\beta$ are broadly consistent with the findings of the H$\alpha$, however they require a more careful treatment of the \ion{Fe}{2} pseudo-continuum emission, which is beyond the scope of this work. In the rest of the paper we adopt the redshift of the source to be $z=0.005839$ as reported for 21-cm estimates by the NASA/IPAC extragalactic database (NED).

\section{Analysis and results}
\label{section:spectra}

In this section, we present the results obtained by modeling the spectra from the different instruments. The X-ray spectral analysis is performed using \textsc{XSPECv12.15.0d} \citep{1996ASPC..101...17A}. We adopt the Cash statistic \citep{Cash79}. All uncertainties are reported at the 90\% confidence level, unless specified otherwise. 

\subsection{X-ray continuum fitting}
\label{section:continuum}

\begin{figure}
  \centering
  \includegraphics[width=\linewidth]{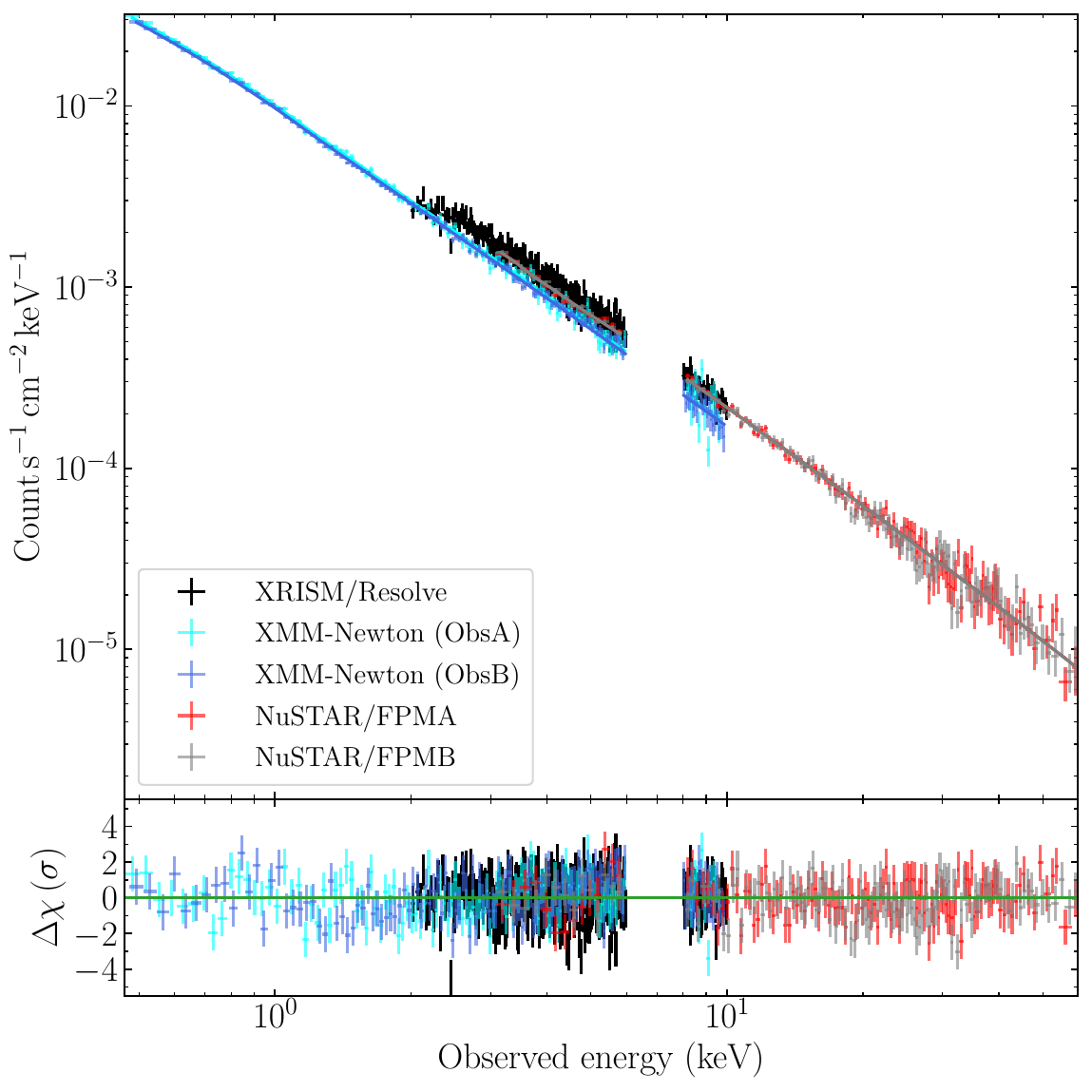}\\
  \includegraphics[width=\linewidth]{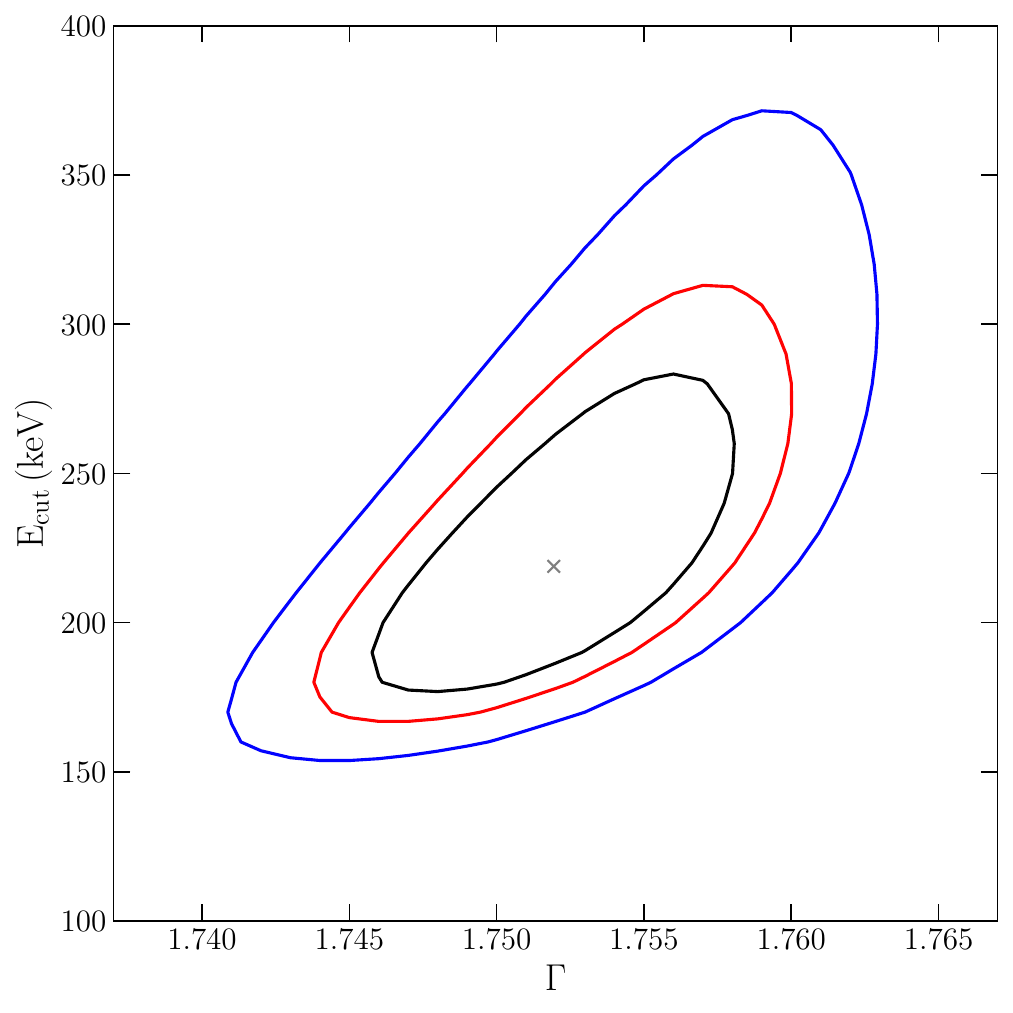}
  \caption{Top panel: XRISM/Resolve (black), XMM-Newton/EPIC-pn and NuSTAR spectra modeled with an absorbed power law with a high-energy cutoff. We also show the corresponding residuals. Bottom panel: The high-energy cutoff versus photon index confidence regions at the 68\%, 95\%, and 99\% levels (black, red, and blue, respectively).} 
  \label{fig:continuum}
\end{figure}

To model the continuum spectrum of the source, we fit the data in the $2-10$\,keV range for XRISM/Resolve, $0.5-10$\,keV range for XMM-Newton/EPIC-pn, and $3-60$\,keV range for NuSTAR/FPMA and FPMB. For all instruments, we ignored the $6-8$\,keV energy range to avoid the complexity introduced by the presence of the Fe emission lines. This exclusion is not expected to affect the continuum fit. We adopt a model that consists of an absorbed power law with a high-energy cutoff. The spectra and the best-fit model are presented in the top panel of Fig.\,\ref{fig:continuum}. The column density of the interstellar absorber is fixed to the Galactic value of $N_{\rm H} = 1.08 \times 10^{20}\,\rm cm^{-2}$ \citep{HI4PI}. We tied all the parameters among the five spectra, including a cross-calibration normalization constant. We fixed this constant to the one for XRISM/Resolve, and let it vary for all the other spectra. The spectra are well fitted with this model, with a $C-{\rm stat/dof} = 2410/2341$. We obtained a best-fit photon index $\Gamma = 1.752 \pm 0.006$, high-energy cutoff $E_{\rm cut} = 220_{-40}^{+60}\,\rm keV$, and a $2-10\,\rm keV$ flux $F_{\rm 2-10\,keV} = 4.74 \pm 0.05 \times 10^{-11}\,\rm erg\,s^{-1}\,cm^{-2}$. The bottom panel of Fig.\,\ref{fig:continuum} shows the $E_{\rm cut}$ versus $\Gamma$ confidence contours. We obtained cross-normalization constants of $1.04\pm0.01$ for NuSTAR FPMA/FPMB, $0.81 \pm 0.01$ and $0.78 \pm 0.01$ for XMM-Newton\footnote{A cross-calibration difference of $\sim 18-20\%$ is already expected between XMM-Newton and NuSTAR as detailed in the XMM-Newton release note \href{https://xmmweb.esac.esa.int/docs/documents/CAL-SRN-0388-1-4.pdf}{XMM-CCF-REL-388}.} ObsA and ObsB, respectively. Assuming a luminosity distance of 22.8\,Mpc, the intrinsic $2-10$\,keV luminosity of the source is $(2.95 \pm 0.03) \times 10^{42}\,\rm erg\,s^{-1}$. Thus, for a bolometric correction $\kappa_{\rm X} = L_{\rm bol}/L_{2-10} = 10$ obtained from the relation provided by \cite{Duras2020} and \cite{Lopez2024}, we find an Eddington ratio $\lambda_{\rm Edd}=L_{\rm bol}/L_{\rm Edd} = 0.001 - 0.01$, given the uncertainty on the black hole mass in this source. During these observations the source is caught in its brightest state since the 1980's \citep{Yan2018}, being $\sim 2$ times brighter than the Chandra/HETG observation. The harder photon index than previously reported is consistent with the `harder when brighter' behavior in this source \citep{Emmanoulopoulos2012}.


\subsection{The Fe line region}
\label{section:Feregion}

\begin{figure*}
  \centering
  \includegraphics[width=\linewidth]{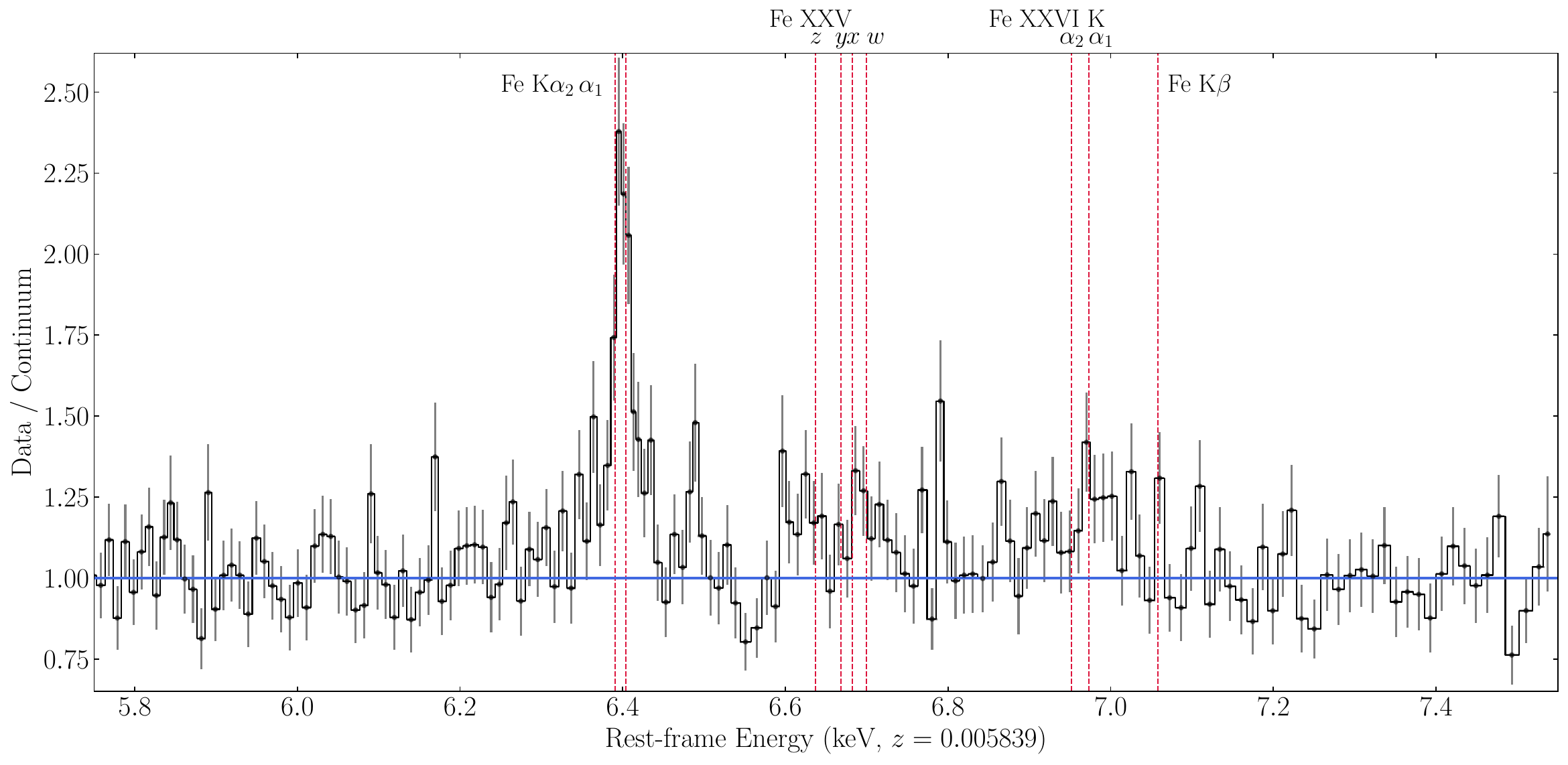}
  \caption{Data to continuum ratio of the XRISM/Resolve spectrum in the Fe lines region in the observed frame. The red dashed lines show the expected energies for the low-ionization Fe\,K$\alpha_{1,2}$, and K$\beta$ lines and the high-ionization \ion{Fe}{25} ($w, x, y, z$) and \ion{Fe}{26}\,K$\alpha_{1,2}$ lines. } 
  \label{fig:XRISM_ratio}
\end{figure*}

In this section, we focus on the fit to the region of the Fe emission lines. We model the XRISM/Resolve spectra only, in the $5.7-7.5\,\rm keV$ range in the observed frame. We assume the continuum follows the best-fit model derived in the previous section. We fix the photon index and the high-energy cutoff to their best-fit values, and we leave the normalization of the power law free to vary. In Fig.\,\ref{fig:XRISM_ratio}, we show the ratio of the XRISM/Resolve data to the continuum. This figure clearly shows an excess in emission centered around the expected energies for the neutral Fe\,K$\alpha_{1,2}$ doublets and a hint of excess at the expected energy for Fe\,K$\beta$ line. We also see an excess in emission at the expected energies for \ion{Fe}{25} and \ion{Fe}{26}. The neutral Fe\,K$\alpha$ line shows a clear, narrow core with a broader, asymmetric base. The \ion{Fe}{25} and \ion{Fe}{26} lines appear at a lower intensity with an apparent broadening. We note that we examined the data for intra-observation variability. However, no strong variability could be confirmed. We found only a $\sim 2\sigma$ hint of variability in the intensity of neutral Fe\,K$\alpha$ emission line. Thus, we analyzed the time-integrated spectrum in the rest of the paper.

\begin{figure*}
  \centering
  \includegraphics[width=0.49\linewidth]{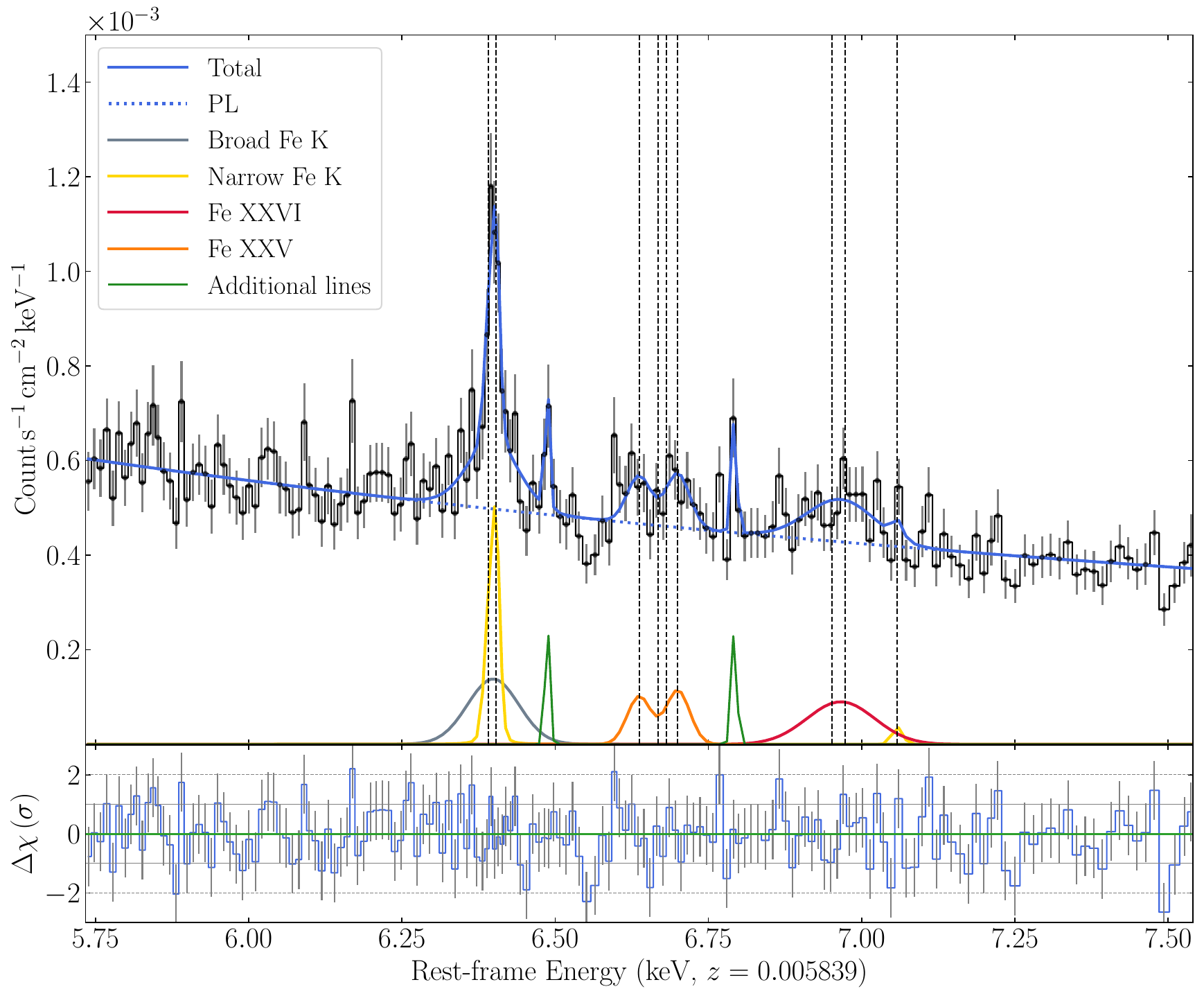}
  \includegraphics[width=0.49\linewidth]{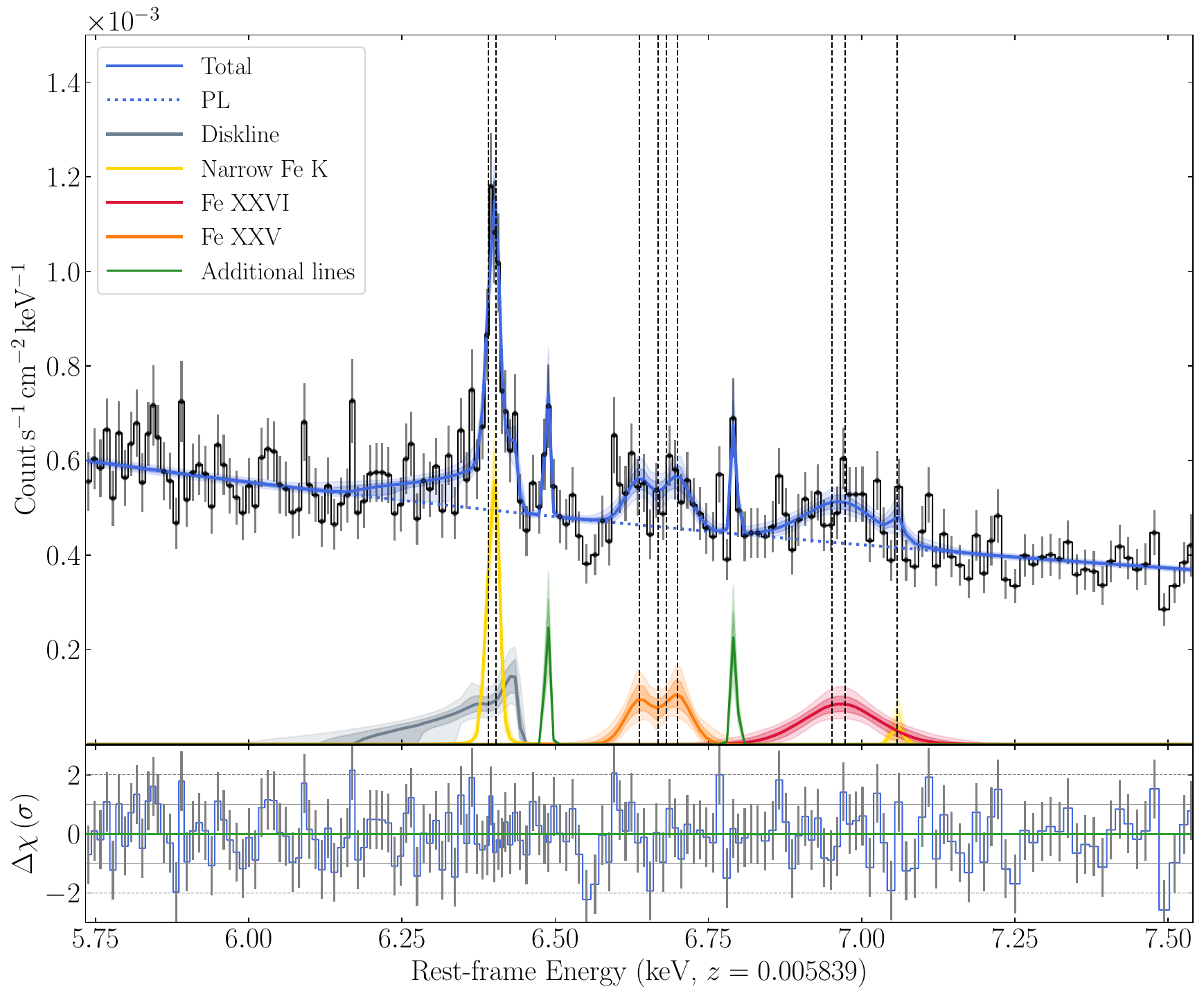}
  \caption{Modeling the XRISM/Resolve spectrum (focusing on the Fe region). In addition to the continuum (blue dotted line), we also show the narrow and broad Fe\,K$\alpha$ lines (yellow and grey, respectively), the \ion{Fe}{25} emission lines (orange), the \ion{Fe}{26} lines (red), and two Gaussian emission lines added at 6.45\,keV and 6.75\,keV (green). The left panel shows the best-fit assuming that the broad Fe\,K$\alpha$ is described with a \textsc{zFeklor} model smoothed with a Gaussian profile. The right panel shows the best-fit assuming that the broad Fe\,K$\alpha$ is described with a \textsc{Diskline} model. The results in the right panel are based on MCMC analysis. The solid lines and the shaded region correspond to the median profile and the 68\% and 95\% credibility envelopes of each spectral component, respectively.} 
  \label{fig:XRISM_spectra_fit}
\end{figure*}

First, we modeled the Fe\,K$\alpha_{1,2}$ by adding the \textsc{ zFeklor} model in \textsc{XSPEC}, which consists of seven Lorentzian approximations to the line profile of Fe\,K fluorescence from neutral material \citep{Holzer1997}, at the rest frame of the source. In addition, we added a Gaussian line at the rest-frame energy of the neutral Fe\,K$\beta$ (7.06\,keV) for completeness, leaving the normalizations of both lines free to vary. We also smoothed the lines with a Gaussian profile using the \textsc{gsmooth} model in \textsc{XSPEC}. This improved the fit by $\Delta C = -188$ for two additional free parameters, resulting in an FWHM of the Fe\,K line of $ 1300\,\kms$ ($\sigma = 11\,\rm eV$). However, additional residuals were still present bluewards and redwards of the Fe\,K line implying the presence of a second broader line. Thus, we added an additional broadened \textsc{gsmooth $\times$ zFeklor} line which improved the fit by $\Delta C= -14$ for two additional free parameters. We obtained an FWHM of $4500_{-2200}^{+5500}\,\kms$ ($\sigma = 41\,\rm eV$) and line intensity of $(1.5 \pm 0.6) \times 10^{-5}\,\rm photon\,s^{-1}\,cm^{-2}$. This reduced the FWHM of the narrow line to 650\,\kms. We then added a series of four Gaussians (\textsc{zvgaussian}) emission lines assuming the rest-frame energies of the \ion{Fe}{25} resonant ($w$, at $E_{\rm rest} = 6.700\,\rm keV$), intercombination ($x$, at $E_{\rm rest} = 6.682\,\rm keV$, and $y$, at $E_{\rm rest} = 6.668\,\rm keV$), and forbidden ($z$, at $E_{\rm rest} = 6.637\,\rm keV$). We assumed the same broadening for all these lines. The intensities of the intercombination lines were low and unconstrained, so we fixed them to zero. The addition of the \ion{Fe}{25} lines improved the fit by $\Delta C = -25 $ for three additional free parameters (the intensities of the $w$ and $z$ lines and the common FWHM). We obtained an FWHM of the \ion{Fe}{25} lines of $2030\,\kms$. Then, we added two Gaussian lines (\textsc{zvgaussian}) at the rest-frame energies of the \ion{Fe}{26}\,K$\alpha_1$ and K$\alpha_2$ lines (6.973\,keV and 6.952\,keV, respectively) assuming an intensity ratio of 2:1. We also assumed a common broadening for the two lines which is independent of that of the \ion{Fe}{25} lines. The addition of these two lines improved the fit by $\Delta C = -21$ for two more free parameters. We note that allowing the energies of all these lines to shift did not result in any significant improvement. All the lines are consistent with the rest frame of the source. Finally, we considered adding two Gaussian lines to account for the narrow excess seen at $\sim 6.45$\,keV and $\sim 6.75$\,keV. We assumed that both lines are unresolved. The addition of these two lines improved the fit by $\Delta C= -15$ and $-14$, respectively, for one additional free parameter each. The observed energies of these two lines are 6.449\,keV and 6.753\,keV, corresponding to rest-frame energies of 6.489\,keV and 6.794\,keV, respectively, assuming that the lines are emitted by \ngc. The possible origin of these lines is discussed in Section\,\ref{sec:gaussians}. In the following, we fix the lines at their best-fit energies, leaving only their corresponding normalizations free. After including all these spectral components, the best-fit statistics obtained are $C/\rm dof = 517/516$. The best-fit model is shown in the left panel of Fig.\,\ref{fig:XRISM_spectra_fit}. The model in \textsc{XSPEC} can be expressed as follows:

\begin{eqnarray}
\textsc{Model}  &=& \textsc{zpowerlaw}  \nonumber \\
                &+& \textsc{GSmooth} \times \textsc{(zFeKlor}_{\rm FeK\alpha} + \textsc{zgauss}_{\rm FeK\beta})_{\rm n} \nonumber \\
                &+& \textsc{Gsmooth} \times \textsc{zFeKlor }_{\rm FeK\alpha,\, b} \nonumber \\ 
                &+&  2\times\textsc{Zgauss}_{\rm Fe XXV} + 2\times\textsc{Zgauss}_{\rm Fe XXVI} \nonumber \\
                &+&   \textsc{gauss}_{\rm 6.45 keV} +  \textsc{gauss}_{\rm 6.75 keV} .
\label{gaussmodel}
\end{eqnarray}

\noindent The $n$ and $b$ subscripts indicate the narrow and the broad low-ionization Fe\,K$\alpha$ lines.


A Gaussian line profile cannot fully account for the apparent asymmetry in the broad Fe\,K$\alpha$ line. To account for this, we replaced $(\textsc{Gsmooth} \times \textsc{zFeKlor }_{\rm FeK\alpha,\, b})$ in Eq.\,(\ref{gaussmodel}) with a \textsc{Diskline} model, which considers line emission from a relativistic accretion disk \citep{Fabian89}. The model in \textsc{xspec} can be expressed as follows:

\begin{eqnarray}
\textsc{Model}  &=& \textsc{zpowerlaw}  \nonumber \\
                &+& \textsc{GSmooth} \times \textsc{(zFeKlor}_{\rm FeK\alpha} + \textsc{zgauss}_{\rm FeK\beta})_{\rm n} \nonumber \\
                &+& \textsc{diskline} \nonumber \\ 
                &+&  2\times\textsc{Zgauss}_{\rm Fe XXV} + 2\times\textsc{Zgauss}_{\rm Fe XXVI} \nonumber \\
                &+&   \textsc{gauss}_{\rm 6.45 keV} +  \textsc{gauss}_{\rm 6.75 keV} .
\label{diskmodel}
\end{eqnarray}

\noindent In addition to the line energy, which we fixed at 6.4\,keV in the rest-frame of the source, the model's parameters are the inner edge of the disk ($R_{\rm in}$), the outer edge of the disk ($R_{\rm out}$), the disk radial emissivity index\footnote{\textsc{Diskline} assumes that the emissivity of the disk follows a power-law profile $\epsilon(r) \propto r^{\tt Betor10 }$, where ${\tt Betor10} = -q$.} ($q$), and the inclination ($i$). The emissivity index could not be constrained when we let it vary ($q > 0.1$). This limit is consistent with the `standard' emissivity index of $3$, expected in the disk at large radii. Indeed, for an optically thick and geometrically thin $\alpha$-disk the thermal energy dissipation per unit surface area is proportional to $r^{-3}$. Also, for a compact central X-ray corona, the irradiation at distant parts of the accretion disk should follow $r^{-3}$ \citep[see e.g.,][]{Reynolds1997}. Thus, we fitted the XRISM/Resolve spectrum by fixing $q$ at $3$. We also fixed the outer radius at $R_{\rm out} = 5000\,\rg$ because the fit was not sensitive to the value of $R_{\rm out}$. This fit resulted in $C/\rm dof = 510/515$, an improvement by $\Delta C = -7$ for one more free parameter compared to a Gaussian (an overall improvement of $\Delta C = -21$ with respect to the fit with no broad line). To test the goodness of the fit, we use the analytic prescription provided by \cite{Kaastra2017} to estimate the expected $C$-stat and its variance based on the observed spectrum. We find an expected $C$-stat of $533 \pm 33$. Thus, our best fit $C$-stat is consistent with the expected value within less than $1 \sigma$. To sample the posterior distribution of the XRISM/Resolve spectral model, we employed \textsc{XSPEC}'s implementation of the Goodman–Weare affine–invariant ensemble sampler, using 64 walkers and $4\times 10^6$ accepted steps per walker. The first $ 10^6$ steps were discarded as burn-in. The convergence tests and uncertainty estimates are detailed in Appendix\,\ref{appendix:MCMC}. 

In addition, we ran a set of simulations detailed in Appendix\,\ref{appendix:line_sig} to assess the significance of the emission lines with relatively low $\Delta C$ used in our model. While the significance of a broad base of the low-ionization Fe\,K$\alpha$ line is strong, we tested the significance of an asymmetric \textsc{diskline} with respect to a symmetric Gaussian-broadened line. We found that given the quality of our data, the \textsc{diskline} is preferred with a 99.1\% confidence ($2.6 \sigma$) . As for the lines at 6.45\,keV and 6.75\,keV, we find that they are both significant with a 99.9\% confidence ($\sim 3.3\sigma$).

\begin{table}[ht]
  \centering
  \begin{tabular}{llll}
  \hline \hline
  Spectral component & Parameter &  & $\Delta C$ \\ \hline 
  narrow Fe\,K$\alpha$ & $\rm FWHM $ 	&	$ 650_{- 220}^{+240}$ & $-188$\\
  & $\rm Norm_{\rm Fe\,K\alpha}$	&	 $14.3_{- 2.8}^{+2.4}$ &\\
  & $\rm Norm_{\rm Fe\,K\beta}$ 	&	 $< 1.6$ &\\
   & $\rm EW $ 	&	 $ 32 \pm 6$ &\\
  \hline 
  
   \textsc{Diskline} & $i (^\circ)$ 	&	 $11.3_{- 0.9}^{+2.2}$ & $-21$ \\
  & $\log (R_{\rm in}/\rg)$ 	&	 $2.0_{- 0.4}^{+0.5}$  &  \\
 &  $\rm Norm$ 	&	 $16_{- 7}^{+6}$  &  \\
 & $\rm EW $ 	&	 $ 26 \pm 18$  & \\\hline
  
  \ion{Fe}{25} & $\rm FWHM$ 	&	 $2350_{-900}^{+1600}$   & $-25$\\
  & $\rm Norm_f$ 	&	 $5.4 \pm 2.6$  \\
  & $\rm Norm_r$	&	 $5.9 \pm 2.4$ \\ 
  & $\rm EW_f$ 	&	 $ 10 \pm 5$ \\
  & $\rm EW_r$ 	&	 $ 11 \pm 5$ \\\hline
  
  \ion{Fe}{26} & $\rm FWHM$ 	&	 $6100_{- 2300}^{+2500}$ & $-21$ \\
 & $\rm Norm_{K\alpha 1}$ 	&	 $8.9\pm 3.1$ \\ 
 & $\rm EW_{K\alpha 1}$ 	&	 $ 20 \pm 6$ \\ \hline 

  Gaussian lines & $\rm Norm_{\rm E=6.45 \,keV}$	&	 $2.5\pm 0.9$ & $-15$  \\
  & $\rm Norm_{\rm E=6.75 \,keV}$	&	 $2.2\pm 0.9$  & $-14$\\ \hline
  

  \end{tabular}
  \caption{Best-fit parameters obtained by fitting the XRISM/Resolve spectrum. The last column shows the improvement in $C$-stat obtained by introducing each of the lines. FWHMs are in units of \kms. Normalizations are in units of $10^{-6}\,\rm photon\,s^{-1}\,cm^{-2}$. EWs are in units of eV.}
  \label{tab:bestfit-xrism}
\end{table}

Our results demonstrate for the first time in \ngc\ that the Fe\,K$\alpha$ is consistent with being composed of two lines, a narrow core with $\rm FWHM = 650_{-220}^{+240}\,\kms$ and a broader line that is consistent with a disk-like emission originating at a distance of $\sim 100\,\rg$ ($\sim 0.03-3 \times 10^{-3}\,\rm pc$, given the uncertainty of the BH mass). Our best-fit model results in a low inclination of $11\degr$ that is consistent with the unobscured properties and the optical classification of this source. In addition, the high resolution of the XRISM/Resolve spectrum allowed us to resolve the forbidden and the resonant lines of \ion{Fe}{25}, which are broadened with $\rm FWHM =  2350_{-900}^{+1600} \,\kms$. In addition, we found an even broader \ion{Fe}{26}\,K$\alpha_{1,2}$ lines with $\rm FWHM = 6100_{- 2300}^{+2500}\,\kms$.

\subsection{The optical spectrum}

\begin{figure}
  \centering
  \includegraphics[width=1\linewidth]{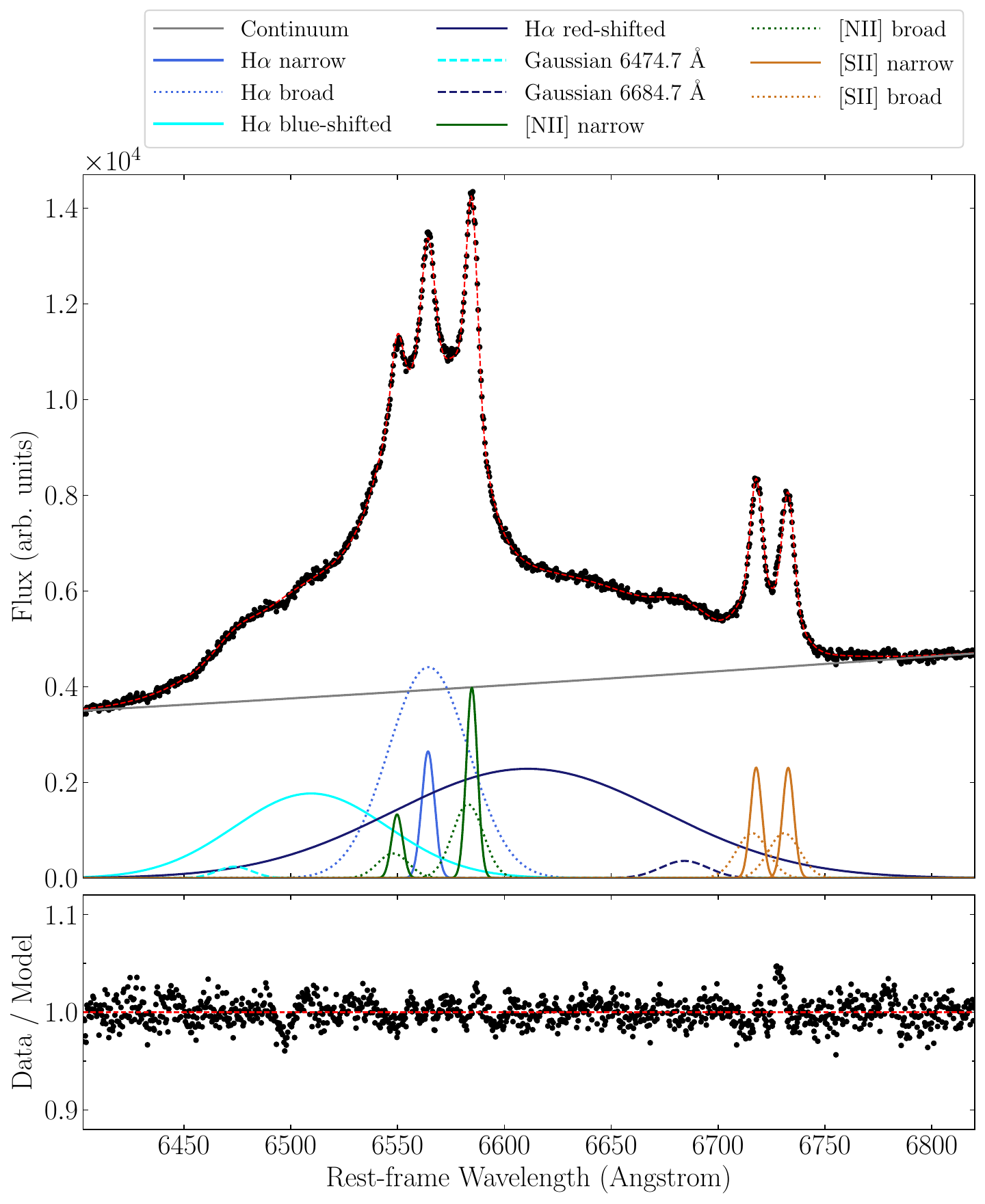}
  \caption{SOAR spectrum of \ngc\ in the H$\alpha$ region. We fit the data assuming a power-law continuum and a series of Gaussian emission lines to model the narrow H$\alpha$, a broad symmetric H$\alpha$ line, two broad red-/blueshifted H$\alpha$ lines, and narrow [\ion{N}{2}]\,$\lambda\lambda6548,6583$ and [\ion{S}{2}]\,$\lambda\lambda6716,6731$ lines. In addition, the data required broad and blueshifted [\ion{N}{2}]\,$\lambda\lambda6548,6583$ and [\ion{S}{2}]\,$\lambda\lambda6716,6731$, as well as a low-amplitude red-/blue-shifted doublet at $\rm \sim 6682\,\AA$ and $\rm \sim 6472\,\AA$, respectively, with an intermediate broadening of $\sim 900-1000\,\kms$.}
  \label{fig:optical_fit}
\end{figure}

\begin{table}
  \centering
  \begin{tabular}{llll}
  \hline \hline
  Line & $\rm \lambda_0$ & $\Delta \lambda$ &FWHM \\ 
     & $\rm (\AA)$ & $\rm (\AA)$ & ($\rm km\,s^{-1}$) \\ \hline

    H$\alpha$ narrow	&		$6564.3	\pm 0.2$	& $1.5 \pm 0.2$	&	$325 \pm 1$		\\
    H$\alpha$ broad	&		$6564.4 \pm 0.4$	&$ 1.6 - 0.4$	&	$2020	^{+4}_{-7}$	\\
    H$\alpha$ blue	&		$6509.9	\pm 0.2$ & $ -52.9 \pm 0.2$	&		$3890	^{+10}_{-20}$	\\
    H$\alpha$ red	&		$6610.9	^{+0.2}_{-0.9}$	& $48.1_{-0.9}^{+0.2}$	&	$6760 	^{+10}_{-20}$	\\
    Gaussian	&		$6474.7	^{+0.3}_{-1.1}$	&	&	$900	^{+1}_{-6}$	\\
    Gaussian	&		$6684.7	^{+0.2}_{-0.9}$	&	&	$1170	^{+3}_{-4}$ \\	
    $[$\ion{N}{2}$]$ narrow & $6549.79 \pm 0.01$	& $1.78 \pm 0.01$	&	$280	\pm 1$ \\
                & $6584.78 \pm 0.01$	& $1.78 \pm 0.01$	&	$280^{\rm tied}$ \\

    $[$\ion{N}{2}$]$ broad	&		$6547.92 \pm 0.01	$ & $-0.08 \pm 0.01$	&	$775 \pm 5$	\\
                  & $6582.92 \pm 0.01$	& $-0.08 \pm 0.01$	&		$775^{\rm tied}$ \\

     $[$\ion{S}{2}$]$ narrow	&		$6717.82 \pm 0.01$	& $1.8\pm 0.01$	&	$280^{\rm tied}$	\\
                    &		$6732.83 \pm 0.01$	& $1.8\pm 0.01$	&	$280^{\rm tied}$	\\

     $[$\ion{S}{2}$]$ broad	&		$6715.91 \pm 0.01$ & $-0.1\pm 0.01$		&		$775^{\rm tied}$ \\
                  &		$6730.91 \pm 0.01$	&  $-0.1\pm 0.01$		& 	$775^{\rm tied}$ \\ \hline
  \end{tabular}
  \caption{Best-fit central rest-frame wavelengths (assuming $z=0.005839$), wavelength shift with respect to the expected value, and FWHM of the emission lines identified in the H$\alpha$ region of the \ngc\ SOAR spectrum.}
  \label{tab:optical_lines}
\end{table}


We model the SOAR spectrum of \ngc, focusing on the H$\alpha$ region assuming a power-law pseudo-continuum. We also include a series of Gaussian lines in emission to account for the narrow H$\alpha$ line, [\ion{N}{2}]\,$\lambda\lambda6548,6583$, and [\ion{S}{2}]\,$\lambda\lambda6716,6731$. We assumed intensity ratios of 1:3 for [\ion{N}{2}]\,$\lambda\lambda6548,6583$ and 1:1 for [\ion{S}{2}]\,$\lambda\lambda6716,6731$. A broad Gaussian line is required by the data with a rest-frame wavelength consistent with that of H$\alpha$. The velocity shift and the width of the narrow lines are linked. We added a couple of broad Gaussian emission lines red- and blue-shifted with respect to the H$\alpha$ rest-frame wavelength that are characteristic of broad double-peaked AGN \citep[see e.g.,][]{Eracleous1994, Schimoia2017}, with a velocity shift of $\sim \pm (2200-2400)\,\kms$. This model could not capture the full complexity of the data, as some residuals in excess could still be visible around the [\ion{N}{2}]\,$\lambda\lambda6548,6583$ and [\ion{S}{2}]\,$\lambda\lambda6716,6731$, which are most likely due to the presence of broad blue wings associated with these lines \citep[see e.g.,][]{Schimoia2017}. By adding broad blue-shifted lines ($\sim -100\,\kms$) with respect to the narrow lines, the residuals flattened around these regions. Finally, we added two broad lines around $\sim 6475\,\rm \AA$ and $6685\,\rm \AA$ with a $\rm FWHM \sim 900-1200\,\kms$, of unknown origin, which are required by the data. The spectrum and the best-fit model are shown in Fig.\,\ref{fig:optical_fit}. We note that the spectrum is not flux calibrated. We also list the wavelengths and the FWHM of each of the lines included in the model in Table\,\ref{tab:optical_lines}. The results of this analysis are consistent with the findings of previous optical spectral studies of this source \citep{Bianchi2008, Schimoia2017}. The wavelengths of the narrow lines are redshifted with respect to the rest-frame of the source by $\sim 80\,\kms$\ with a FWHM of $\sim 300\,\kms$. We obtained an FWHM for the symmetric broad H$\alpha$, thought to originate from the BLR, of $2020_{-7}^{+4}\,\rm \kms$.

The XRISM/Resolve spectrum of \ngc\ revealed for the first time the presence of an asymmetric broad Fe\,K$\alpha$ line. We modeled this line assuming a relativistically broadened disk-like feature with an inner radius of $\sim 100\,\rg$ viewed at an inclination of $\sim 11\degr$. This feature resembles the double-peaked broad emission line seen in H$\alpha$. In fact, various works used a disk-like profile or a variation of it to model optical double-peaked broad emission lines in AGN \citep[see e.g.,][]{Chen1989a, Chen1989b, Storchi-Bergmann1993, Eracleous2003, Strateva2003, Bianchi2019, Bianchi2022, Ward2024, Ward2025}. Such features are commonly modeled using disk-like profiles and relatively flat emissivity indices ($q \sim 0$ to 2) that is smoothed with a Gaussian profile. This smoothing accounts for the local turbulence of the emitting gas, parameterized via a velocity dispersion ($\sigma_{\rm turb}$). Various authors adopt a variety of emissivity profiles. However, the common feature in these works is that the emissivity index is flat in the inner regions and then drops in a faster way at larger radii. After subtracting the continuum as well as all the best-fit lines except the broad red- and blue-shifted lines, we modeled the double-peaked H$\alpha$ with a relativistic \textsc{Diskline} model \citep{Fabian89} modified by a Gaussian smoothing to account for local turbulence. This model is chosen for consistency with the one adopted for the Fe\,K$\alpha$ line, which facilitates the comparison between the two lines. The double-peaked profile and the best-fit model are shown in the left panel of Fig.\,\ref{fig:DPB}. We obtained the best fit with an inclination $i=18.8\degr \pm 0.3\degr$, an inner radius $\log (R_{\rm in}/\rg) = 2.32 \pm 0.05$, an outer radius $\log (R_{\rm out}/\rg) = 3.16 \pm 0.05$, an emissivity index $q=1.6_{-0.3}^{+0.2}$, and a local turbulent broadening $\sigma_{\rm turb} = 1420\pm 50\,\kms$. This turbulence requires a moderately high scale height, consistent with a flatter $q$ value, as expected in low/intermediate accretion rate sources \citep[e.g.,][]{Elitzur2014}. 


\section{Discussion}
\label{sec:discussion}

\subsection{The inner edge of the disk}


\begin{figure*}
  \centering
  \includegraphics[width=0.45\linewidth]{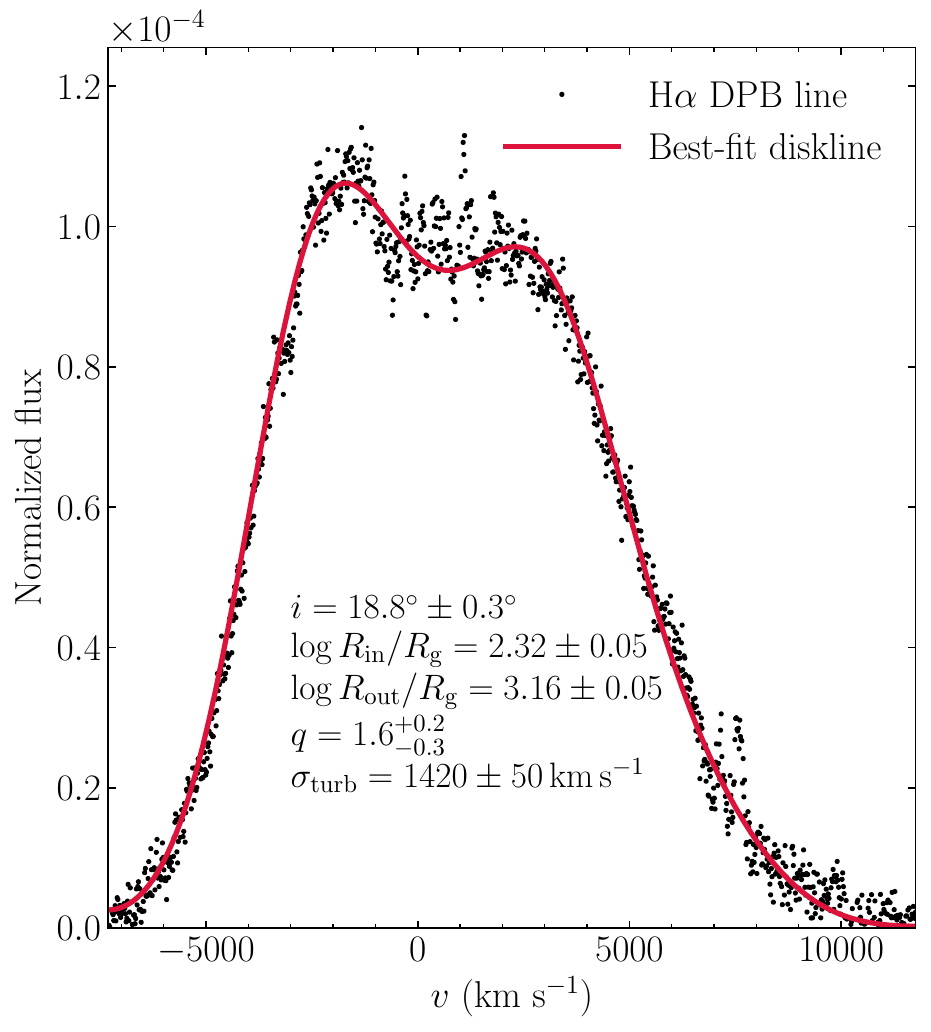}
  \includegraphics[width=0.5\linewidth]{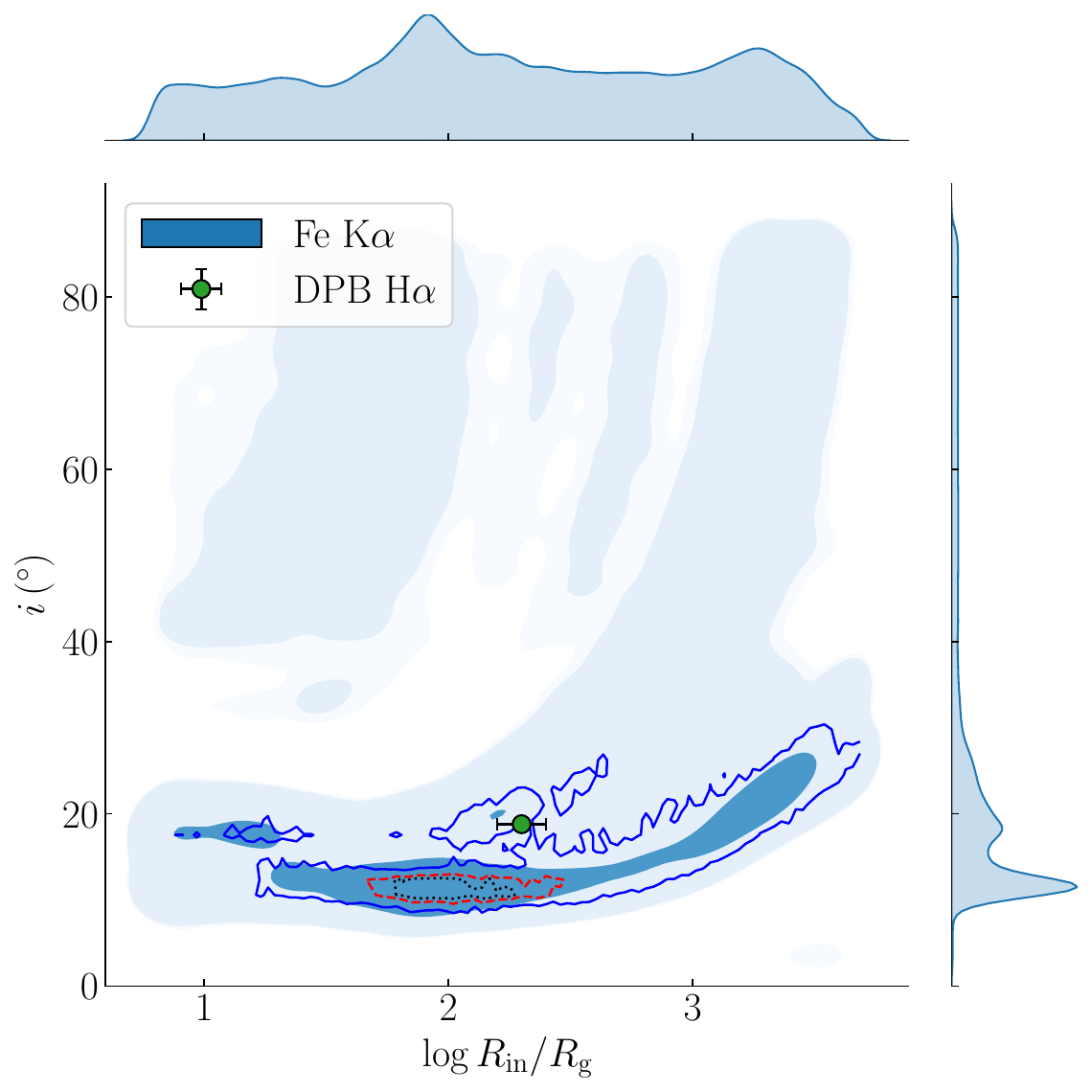}
  \caption{Left: The broad double-peaked H$\alpha$ profile of \ngc\ in velocity space modeled by a \textsc{diskline} with an additional Gaussian smoothing. The negative/positive sign corresponds to blue-/redshift with respect to 6562.8\,\AA.  Right: Confidence contours of the disk inclination versus inner radius obtained by modeling the broad Fe\,K$\alpha$ line in the XRISM/Resolve spectrum. Shaded regions enclose the 68\%, 90\%, 95\% posterior probability. The solid contours correspond to the contours derived from a grid of $\Delta C = 2.3, 4.6,$ and 10 (black, red, and blue, respectively) which would correspond to 68\%, 90\%, and 99.5\% confidence level in the $\chi^2$ approximation for two parameters of interest. The green circle shows the best-fit obtained by modeling the broad double-peaked H$\alpha$ profile on the left. }
  \label{fig:DPB}
\end{figure*}


In the right panel of Fig.\,\ref{fig:DPB}, we show the marginal and joint probability-density plots of the inclination and inner radius obtained by modeling the broad asymmetric Fe\,K$\alpha$. The blue, filled contours represent the posterior probability density obtained from the XRISM/Resolve MCMC run. Within the main, high-probability island the two representations overlap closely, illustrating that the two measures are fully consistent in the region of interest. The faint, elongated ridge that stretches towards higher inclinations and larger inner radii in the plane is a genuine mode explored by the sampler, but it carries very little posterior mass. Integrating the kernel density estimation (KDE) gives a probability of $\sim 2\times10^{-3}$. Consistent with that, the smallest $C$-stat encountered anywhere along the ridge still lies at $ \Delta C \simeq 13$ above the global minimum. We therefore regard the ridge as a statistically insignificant tail of the posterior and conduct all the quantitative inference with respect to the dominant peak. 

We also show in this panel the best-fit inclination and inner radius derived from fitting the broad double-peaked H$\alpha$. The inner radii derived from both lines are fully consistent. The inclination derived from both lines agrees within $3\,\sigma$, although the H$\alpha$ profile requires a slightly larger inclination. The emissivity index of the Fe\,K$\alpha$ could not be constrained ($q > 0.1$) and is found to be broadly compatible with the standard value of $q=3$, the emissivity index of the H$\alpha$ is significantly flatter. We attempted to fit the broad Fe\,K$\alpha$ fixing all the parameters to the best-fit values obtained by modeling the H$\alpha$ line. This worsens the fit by $\Delta C = 13$. This difference could have a physical origin. The double-peaked Balmer lines are thought to originate from the atmosphere of the accretion disk and/or the base of weak disk winds typically seen at low accretion rates \citep[see e.g.,][and references therein]{Eracleous2009}. It has been demonstrated that the double-peaked broad Balmer lines are preferentially produced in optically thin media \citep[$\tau \lesssim 1$; see e.g.,][]{Murray1997, Flohic2012, Chajet201}. As the optical depth increases, the two peaks become closer and resemble a single-peaked emission line. The Fe\,K$\alpha$, instead, is thought to be produced at higher optical depths ($\tau \sim 1-2$). This could indicate that the two lines are produced at the same radial distance from the central engine, though in different locations along the vertical extension of the disk. The H$\alpha$ could be produced in the outer atmosphere, and the Fe\,K$\alpha$ deeper in the disk. In addition, the turbulent velocity of the H$\alpha$ line ($\sigma_{\rm turb} \sim 1400\,\kms$) also supports the idea that this line is produced in a rather turbulent medium, consistent with the outer atmosphere of the accretion disk. These physical considerations can consequently explain the marginal differences we see in inclination and emissivity index. This discrepancy could also arise from some model-dependent degeneracies between the various parameters of the model. However, given the quality of the data, the parameters of the two lines are consistent within $\sim 3\,\sigma$. 

\cite{Schimoia2017} presented a monitoring campaign of the H$\alpha$ emission line in this source over $\sim 2$\,years. They showed that the profile of the double-peaked feature is variable, and attributed this behavior to a precessing spiral arm in the accretion disk with a variable intensity. They also assumed a broken power-law emissivity profile where the inner emissivity index between $R_{\rm in}$ and a break radius ($R_{\rm q}$) is fixed at $-0.2$, and the outer emissivity index above $R_{\rm q}$ is fixed at $3$. They found $R_{\rm in} = 300 \pm 60 \,\rg$, $R_{\rm q}$ varying between 1800\,\rg\ and 3000\,\rg, $i = 47\degr \pm 2\degr$. The inner and outer (break) disk radii found by \cite{Schimoia2017} are broadly consistent with the values we find in this work. The difference between their best-fit inclination and emissivity index and ours could be perceived as significant. This difference can be explained by degeneracies in the model used by \cite{Schimoia2017}. In fact, the \cite{Schimoia2017} model has ten parameters as opposed to only four in our model. We performed a detailed comparison between the two models confirming that the line profile presented in Fig.\,\ref{fig:DPB} could be reproduced with the model by \cite{Schimoia2017} assuming $i = 47\degr$.

We note that a FWHM of the broad Fe\,K$\alpha$ line is consistent with the FWHM of the symmetric core of the broad H$\alpha$ line at $\sim 2.7\sigma$. This does not completely rule out an origin of the broad Fe\,K$\alpha$ that is consistent with the BLR similar to what is inferred in other sources \citep[e.g.,][]{Bogensberger2025}. However, the fact that the a disk origin of the Fe\,K$\alpha$ is statistically preferred, and the properties of the line are in agreement with the double-peaked broad H$\alpha$ line strongly suggest that the  broad Fe\,K$\alpha$ line may be originating from regions closer to the SMBH. In any case, it is likely that a wide range of radii, extending up to the optical BLR, could contribute to the broad Fe\,K$\alpha$ line given a certain emissivity profile. In addition, the BLR itself may also have disk-like shape or be closely associated with the accretion disk as shown by results from BLR reverberation mapping studies\citep[e.g.,][]{PozoNunez2013, Pancoast2014, Grier2017, Williams2020}.

\subsection{The narrow Fe\,K$\alpha$ line}

\begin{figure}
  \centering
  \includegraphics[width=\linewidth]{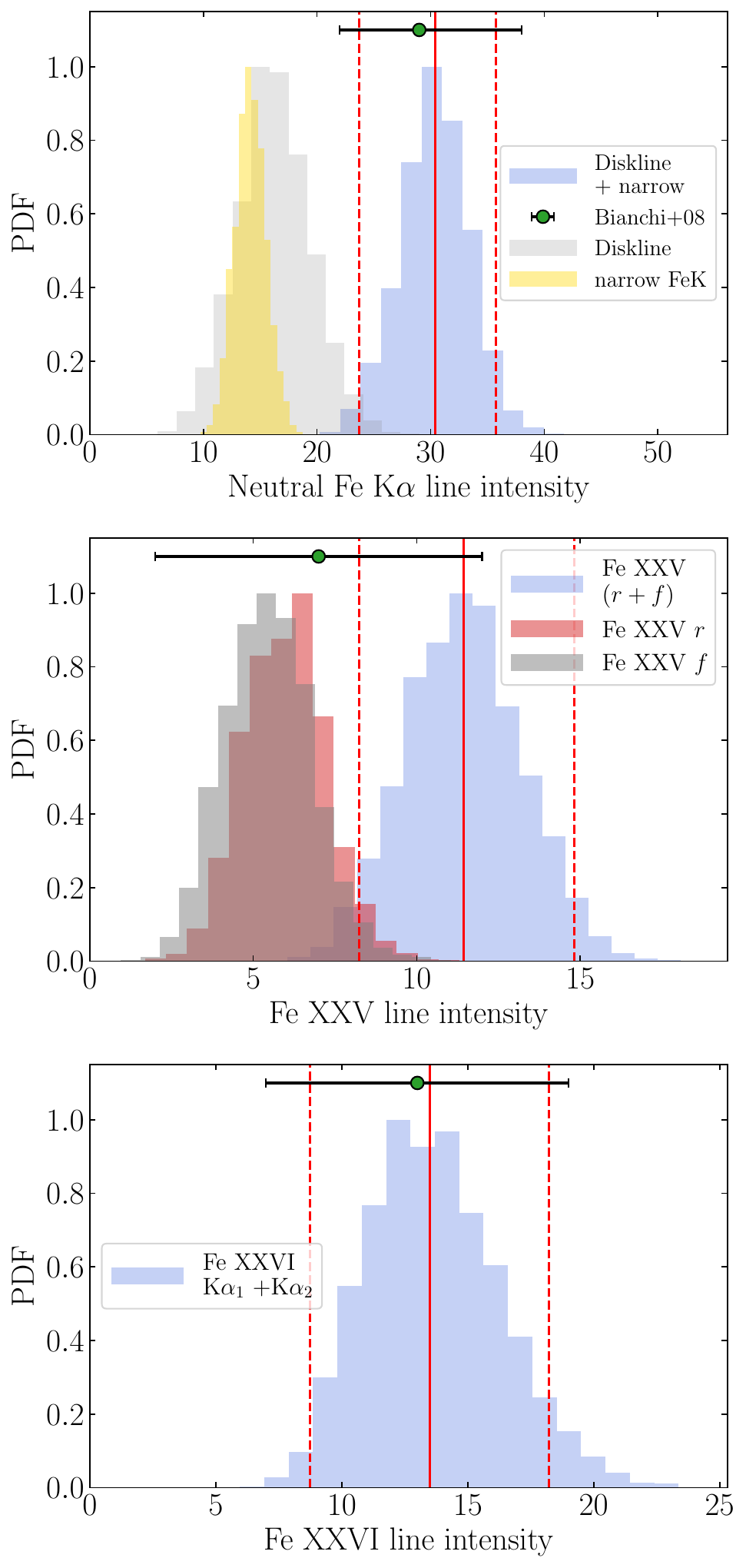}
  \caption{Posteriors of line fluxes obtained from the best-fit to the XRISM/Resolve data (in units of $10^{-6}\,\rm photon\,s^{-1}\,cm^{-2}$). Top: line fluxes of the narrow, \textsc{diskline} Fe\,K$\alpha$, and their sum (yellow, gray, and blue, respectively). Middle: line fluxes of the forbidden and resonant \ion{Fe}{25} lines, and their sum (grey, red, and blue, respectively). Bottom: line fluxes of the \ion{Fe}{26}\,K$\alpha_{1,2}$. The solid and dashed lines show the median and the 90\% confidence range of the distributions, respectively. The green circles show the best-fit flux lines of the corresponding lines from Chandra/HETG reported by \cite{Bianchi2008}.   }
  \label{fig:line_intensities}
\end{figure}

The XRISM/Resolve spectrum shows a clear narrow Fe\,K$\alpha$ line with a $\rm FWHM = 650_{-220}^{+240}\,\kms$. Assuming that the emitting gas follows an ordered rotation motion, the relation between the rotation velocity and the FWHM can be written as  ${\rm FWHM} = 2 v_{\rm rot} \sin i $. The distance of the emitting material can then be written as $R = GM_{\rm BH}/v_{\rm rot}^2$. For an inclination of $11\degr-19\degr$, a BH mass of $0.2-2.4\times 10^8\,\ms$, and also accounting for the uncertainty on the FWHM of the line, we obtain a radius of the narrow Fe\,K$\alpha$ emitting region of $\sim (1.6-18)\times 10^4\,\rg$ ($\sim 0.01-2.1\,\rm pc$). This large uncertainty is mainly driven by the uncertainty on the SMBH mass. We fitted the narrow Fe\,K$\alpha$ with a \textsc{zFeKlor} model blurred with an \textsc{rdblur} \citep{Fabian89} kernel (equivalent to a \textsc{diskline} model but accounting for the fact that Fe\,K$\alpha$ is a doublet, which matters at such low velocities). We linked the inclination of this component to that of the broader Fe\,K$\alpha$ line and left the inner edge free to vary. We obtained $R_{\rm in} = 1.4_{-0.7}^{+2.0} \times 10^{4}\,\rm \rg$, consistent with our previous estimate, although with a smaller uncertainty. Considering the uncertainty on the BH mass, we obtain a physical radius of $0.01-0.4\,\rm pc$. 

Following the estimate by \cite{Nenkova2008}, we calculate the dust sublimation radius as a function of the bolometric luminosity and the dust temperature, as follows:
\begin{equation}
  R_{\rm sub} = 0.4 \left( \frac{L_{\rm bol}}{10^{45}\,\rm erg\,s^{-1}} \right)^{1/2} \left( \frac{1500\,\rm K}{T_{\rm sub}} \right)^{2.6}\,\rm pc.
\end{equation}

\noindent For $L_{\rm bol} = 3 \times 10^{43}\,\rm erg\,s^{-1}$ (see Sect.\,\ref{section:continuum}) and a dust sublimation temperature between 1000\,K and 1800\,K, we obtain a dust sublimation distance of $0.04-0.2$\,pc, which is fully consistent with the distance of the Fe\,K$\alpha$ emitting region. The presence of a narrow Fe\,K$\alpha$ consistent with the dust sublimation radius could be an indication of a dusty torus. However, it is worth noting that the hard X-ray spectrum of \ngc\ from our analysis (Section\,\ref{section:continuum}) and a previous NuSTAR observation of the source \citep{Ursini2015} does not show any indication of a Compton hump, which typically indicates reprocessing from cold material. This absence of a strong Compton hump indicative of either a low column density of this material, and/or of a small solid angle subtended by the torus. This also applies to the relativistically broadened Fe line. A detailed analysis of the broadband spectra of the source is deferred to a future publication. 

We used Eq. (5) of \cite{Murphy2009} that relates the Fe\,K$\alpha$ line EW to the torus covering fraction ($f_{\rm cov}=\Delta \Omega/4\pi$) and the column density which can be approximated as follows:

\begin{equation}\label{eq:EW}
    {\rm EW_{Fe\,K\alpha}} \simeq 580 \times f_{\rm cov} \times \frac{N_{\rm H}}{10^{24}\, \rm cm^{-2}}\, \rm eV,
\end{equation}
\noindent assuming a photon index $\Gamma = 1.75$. We estimate the covering fraction of the torus to be between $4-7\%$ for $N_{\rm H}=10^{24}\,\rm cm^{-2}$ and between $45-65\%$ for $N_{\rm H}=10^{23}\,\rm cm^{-2}$, given the line EW reported in Table\,\ref{tab:bestfit-xrism}. \cite{Ursini2015} estimated a torus column density $N_{\rm H} = 5_{-1.6}^{+2.0}\times 10^{23} \,\rm cm^{-2}$ which results in a covering fraction of  $\sim 6-20\%$ considering the uncertainties on the $\rm EW_{Fe\,K\alpha}$ and $N_{\rm H}$. In addition, the reflection fraction ($R$) from neutral, optically thick slab, can be derived from the Fe\,K$\alpha$ EW as $R\simeq \rm EW_{Fe\,K\alpha} /(150\,\rm eV)$ \citep[e.g.,][]{Geo91}. This results in $R = 0.17 \pm 0.12$. This value is consistent with the absence of a strong Compton hump in the NuSTAR spectrum, and with the 3$\sigma$ upper limit of $R<0.2$ derived by \cite{Ursini2015}.
 
The top panel of Fig.\,\ref{fig:line_intensities} shows the 1D posterior distribution of the best-fitted line intensities of the narrow and the broad Fe\,K$\alpha$ lines. For comparison, we also show the line intensity obtained by \cite{Bianchi2008}, which is consistent with the sum of the intensities of the narrow and the broad lines. In fact, \cite{Bianchi2008} modeled the Fe\,K$\alpha$ from Chandra/HETG assuming a single Gaussian line with a $\rm FWHM = 2400\,\kms$. The low spectral resolution of Chandra/HETG compared to XRISM/Resolve did not allow the authors to distinguish the two lines. Thus, the reported line intensity includes the contribution of both features.

In addition, we found only a weak evidence of an Fe\,K$\beta$ line. The 3$\sigma$ upper limit on the Fe\,K$\beta$/Fe\,K$\alpha$ ratio is 0.2. This is consistent with the 90\% upper limit of 0.18 reported by \cite{Bianchi2008}, which is also consistent with the expectation from neutral iron \citep[see Section\,3.3 in][for example]{Molendi2003}.

We note that we also tested whether the asymmetry seen in the Fe\,K$\alpha$ line could be caused by a Compton shoulder associated with the neutral emission from distant material, by testing reflection models. But, this test did not provide a satisfactory statistical and physical explanation of the data. On the one hand, the model left considerable residuals, mainly failing to explain the blue part of the Fe line. In addition, any significant Compton hump would require relatively large column densities which would then imply strong Compton hump that is in tension with the NuSTAR hard X-rays spectra. In order to be able to account for this, large values of Fe abundance may be required (more than five times the solar abundance) to suppress the relative intensity of the Compton hump with respect to the Fe line. In addition, the fact that the properties inferred from modeling the Fe\,K$\alpha$ line profile with disk emission agree with the properties inferred from the DPB H$\alpha$ profile further support the hypothesis that the Fe K$\alpha$ originates from the accretion disk.

\subsection{The origin of the \ion{Fe}{25} and \ion{Fe}{26} lines}

\begin{figure}
  \centering
  \includegraphics[width=\linewidth]{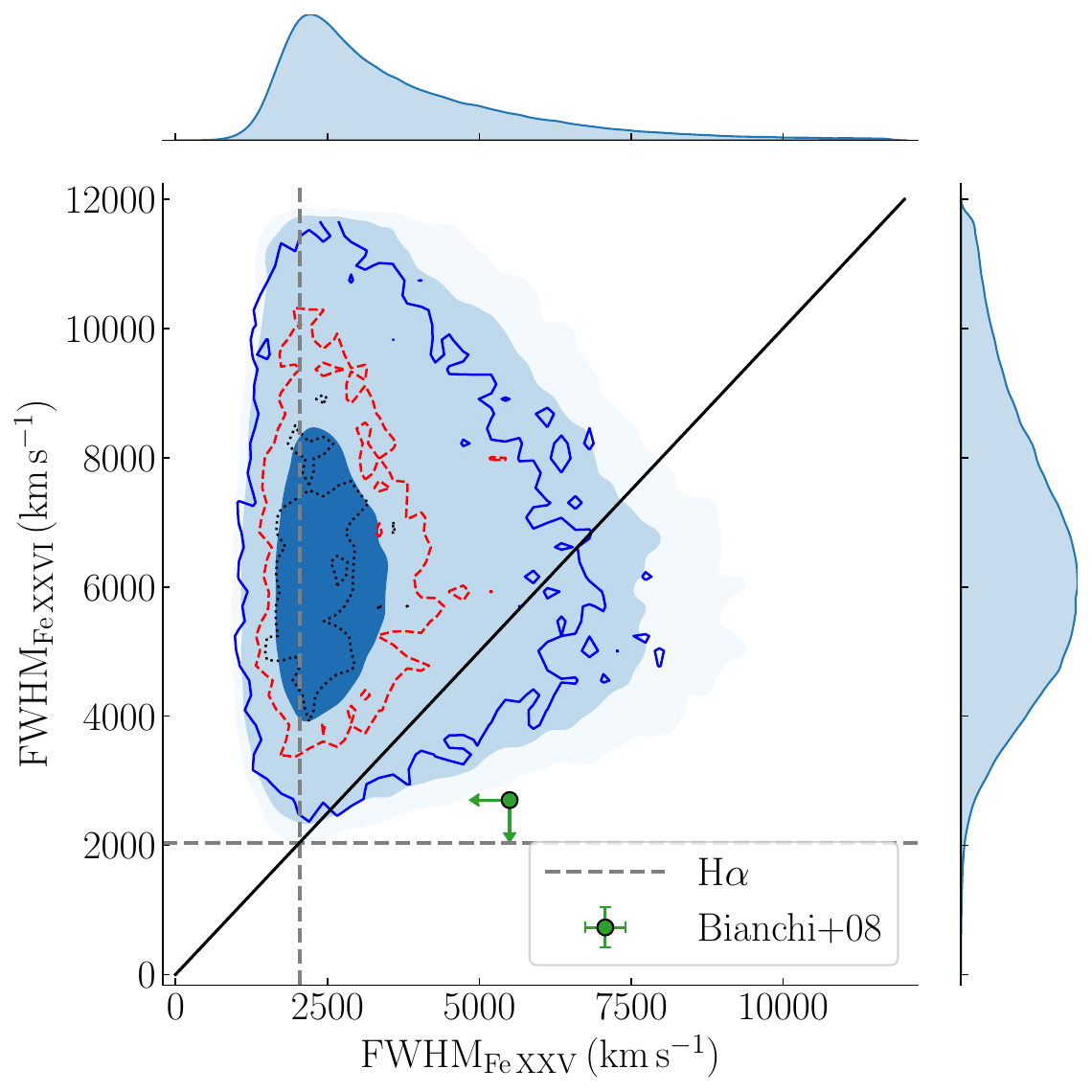}
  \caption{Confidence contours of FWHM of \ion{Fe}{25} and \ion{Fe}{26}. The solid contours correspond to the contours derived from a grid of $\Delta C = 2.3, 4.6,$ and 10 (black, red, and blue, respectively). The green circle shows the corresponding limits reported by \cite{Bianchi2008}. The dashed gray lines show the FWHM of the symmetric broad H$\alpha$ line. The solid black line shows the 1:1 relation.}
  \label{fig:Fexxv_fwhm}
\end{figure}

\begin{figure}
  \centering
  \includegraphics[width=\linewidth]{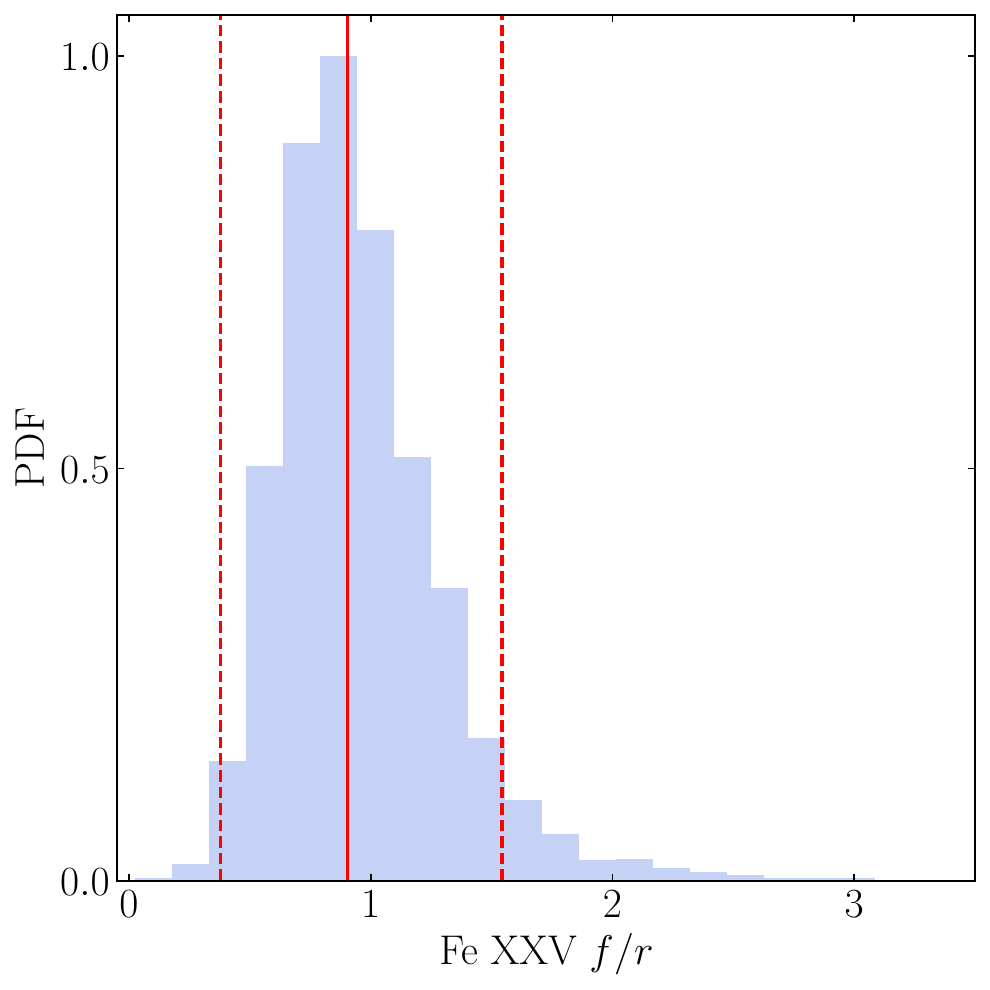} 
  
  \caption{The posterior distribution of the \ion{Fe}{25} forbidden to resonant line ratio. The red solid and dashed lines correspond to the median and 90\% confidence region. }
  \label{fig:Fexxvratio}
\end{figure}

We show in Section\,\ref{section:spectra} that the XRISM/Resolve data required a significant broadening of the highly-ionized \ion{Fe}{25} and \ion{Fe}{26} emission lines. Fig.\,\ref{fig:Fexxv_fwhm} shows the 2D confidence contours of the FWHMs of the two lines. \ion{Fe}{26} requires a broader profile. However, the broadening of the two lines is consistent within $3\,\sigma$. The broadening of the \ion{Fe}{25} lines is consistent with the symmetric core of the H$\alpha$ profile ($\rm FWHM = 2041_{-6}^{+7}\,\kms$). Similar to the Fe\,K$\alpha$ line, we estimated the distance of the emitting material from the BLR to the central SMBH. This results in a distance of the optical BLR of $\sim 8000-31000\,\rg$ ($\sim 0.003-0.1\,\rm pc$). These values agree with the variability timescales of $\sim 7-21\,\rm days$ of the symmetric broad H$\alpha$ reported by \cite{Schimoia2017}. If we assume that the broadening of the \ion{Fe}{25} and \ion{Fe}{26} lines is due to the motion of the orbiting gas around the BH, the \ion{Fe}{25} emitting region is consistent with the optical BLR. While the \ion{Fe}{26} line comes from  closer to the SMBH at a distance of $\sim 178-2500\,\rg$ ($\sim 0.2-30\times 10^{-3}\,\rm pc$). We show in Fig.\,\ref{fig:Fexxvratio} the flux ratio of the forbidden over resonant \ion{Fe}{25} lines. We find a median value of $0.9_{-0.5}^{+0.6}$, which cannot be conclusive about the origin of the lines. Testing various photoionized versus collisionally ionized scenarios will be presented in a forthcoming publication (K. Murakami et al., in prep.).

\cite{Bianchi2008} modeled the \ion{Fe}{25} and \ion{Fe}{26} emission lines in the Chandra/HETG spectrum using single Gaussian lines, consistent with the resonant line (6.7\,keV) and the weighted mean of the \ion{Fe}{26} doublet (6.966\,keV), respectively. Those authors report a hint of blueshift in these lines by $ 21_{-16}^{+10}\,\rm eV$  (i.e., $\sim 1000\,\kms$). Given the fact that only a resonant \ion{Fe}{25} line was detected, and the potential blueshift, the authors suggested that these lines could be originating from starburst activity in the host galaxy. Similar blueshift hints have been reported by \cite{Shi2022}, who attributed this to hot winds from the accretion flow \citep[see also][]{Shi2024}. However, our results do not show any shift in the line energies with respect to the rest frame of the source in the XRISM/Resolve spectra. This could indicate an intrinsic change in the state of the accretion flow as the source transitioned to higher fluxes (a factor of $\sim 2$), or perhaps statistical limitations in the Chandra/HETG data. We show in the bottom two panels of Fig.\,\ref{fig:line_intensities} the 1D distributions of the best-fitted intensities of the \ion{Fe}{25} and \ion{Fe}{26} lines, respectively. We also show the corresponding intensities reported by \cite{Bianchi2008}. For the \ion{Fe}{25}, we show the individual histograms of the resonant and forbidden lines, as well as the sum of the two lines. Our estimate of the resonant line intensity is consistent with the one reported by \cite{Bianchi2008}. We note that \cite{Bianchi2008} did not identify the presence of a forbidden line. This suggests that the properties of the lines may have changed, likely responding to the change in the flux of the source. As for the \ion{Fe}{26} intensity, presented in the bottom panel of Fig.\,\ref{fig:line_intensities}, we show the sum of the doublet from our analysis (assuming an intensity ratio of 2:1) which is consistent with the value reported by \cite{Bianchi2008}.

To further explore the possibility of long-term variability in \ion{Fe}{25} and \ion{Fe}{26}, we show in Fig.\,\ref{fig:Fexxv_fwhm} the upper limits on the FWHM of each of the lines reported by \cite{Bianchi2008}. The upper limit on the \ion{Fe}{25} FWHM of 5500\,\kms\ is consistent with our estimate of the line width in XRISM/Resolve. However, the tight upper limit reported by \cite{Bianchi2008} on the FWHM of the \ion{Fe}{26} line ($<2700\,\kms$) is lower by a factor of $\sim 3$ compared to the value we find from the XRISM/Resolve spectrum. This further supports a scenario in which the highly ionized Fe lines are responding to the change in the intrinsic luminosity of the AGN (twice that of the Chandra/HETG flux), favoring photoionization as the dominant excitation mechanism. Furthermore, the extreme broadening of the lines, notably the \ion{Fe}{26} FWHM exceeding 3800\,\kms at 90\% confidence level, is a strong signature of an AGN origin. It is hard to produce a comparable broadening in the potential of the host galaxy. The detection of such broad lines in optical spectra is usually taken as a strong AGN signature. Finally, it is hard for processes other than AGN activity to produce photons that are energetic enough to ionize iron to \ion{Fe}{26}. In fact, the detection of even much lower ionization states of iron (i.e., [\ion{Fe}{10}) is commonly used as an indicator of AGN in dwarf galaxies \citep[see e.g.,][and references therein]{Molina2021, Reefe2023}. The combination of these three arguments makes a very strong case that these lines cannot come from the host galaxy. Further deep observations of the source at a different flux state will be essential to confirm the variability in these lines and identify their origin.

\subsection{Additional features}
\label{sec:gaussians}

\begin{figure}
  \centering
  \includegraphics[width=\linewidth]{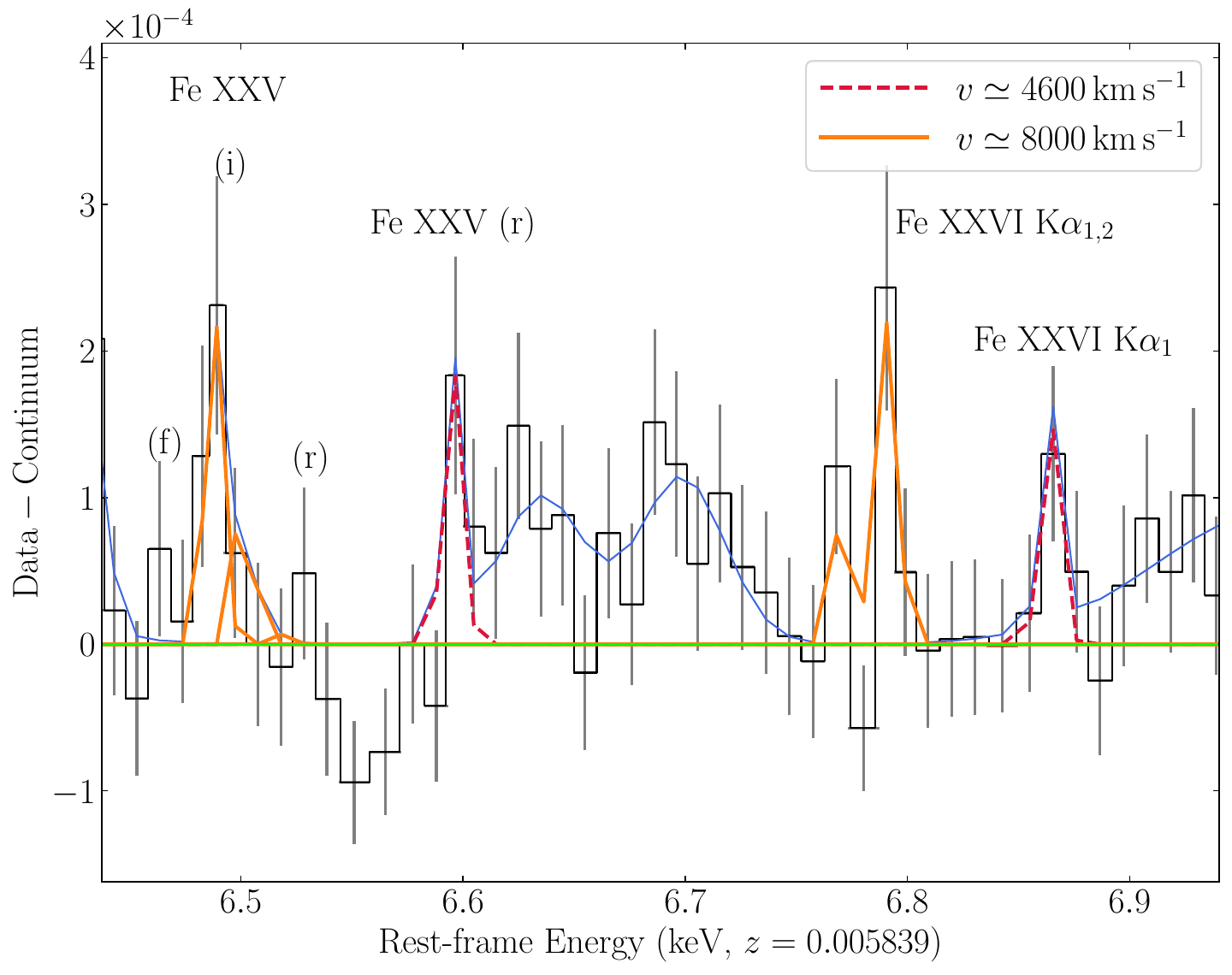}
  \caption{Residual of XRISM/Resolve data after subtracting the power-law continuum (in units of ${\rm Count\,s^{-1}\,cm^{-2}\,keV^{-1}}$), focusing on the observed $6.4-6.9$\,keV range. The blue line shows the total model fitting the emission lines in the residual. The orange solid and red dashed lines show the Gaussian lines  assuming redshifted velocity  of $\sim 8000\,\kms$ and $\sim 4600\,\kms$, respectively, assuming \ion{Fe}{25} and \ion{Fe}{26}.}
  \label{fig:redshifted_lines}
\end{figure}

In addition to the features discussed earlier, the XRISM/Resolve spectrum shows the presence of two prominent emission lines at 6.45\,keV and 6.75\,keV of unknown origin. We modeled these features assuming Gaussian emission line profiles with an unresolved width fixed at 100\,\kms\ (comparable to the XRISM/Resolve energy resolution). We tested the hypothesis that these lines are produced by redshifted \ion{Fe}{25} and \ion{Fe}{26} lines. To do so, we replaced the single Gaussian line by a quadruplet for \ion{Fe}{25} with free intensities, and a doublet for \ion{Fe}{26} with a fixed intensity ratio of 2:1. Under these assumptions, we obtained redshift velocities that are broadly consistent for \ion{Fe}{25} and \ion{Fe}{26}, estimated to be $8040_{-70}^{+100} \,\kms$ and $7730_{-70}^{+80}\,\kms$, respectively. This model also requires the \ion{Fe}{25} emission to be dominated by the intercombination line with forbidden and resonant lines consistent with zero. Investigating the origins of these lines and testing various scenarios goes beyond the scope of this paper. Alternatively, if we assume that these two lines correspond to blueshifted low-ionization Fe\,K$\alpha$ and resonant \ion{Fe}{25} line, this would result in a similar blueshift for both lines of the order of $-4000\,\kms$.

\begin{figure*}
  \centering
  \includegraphics[width=0.9\linewidth]{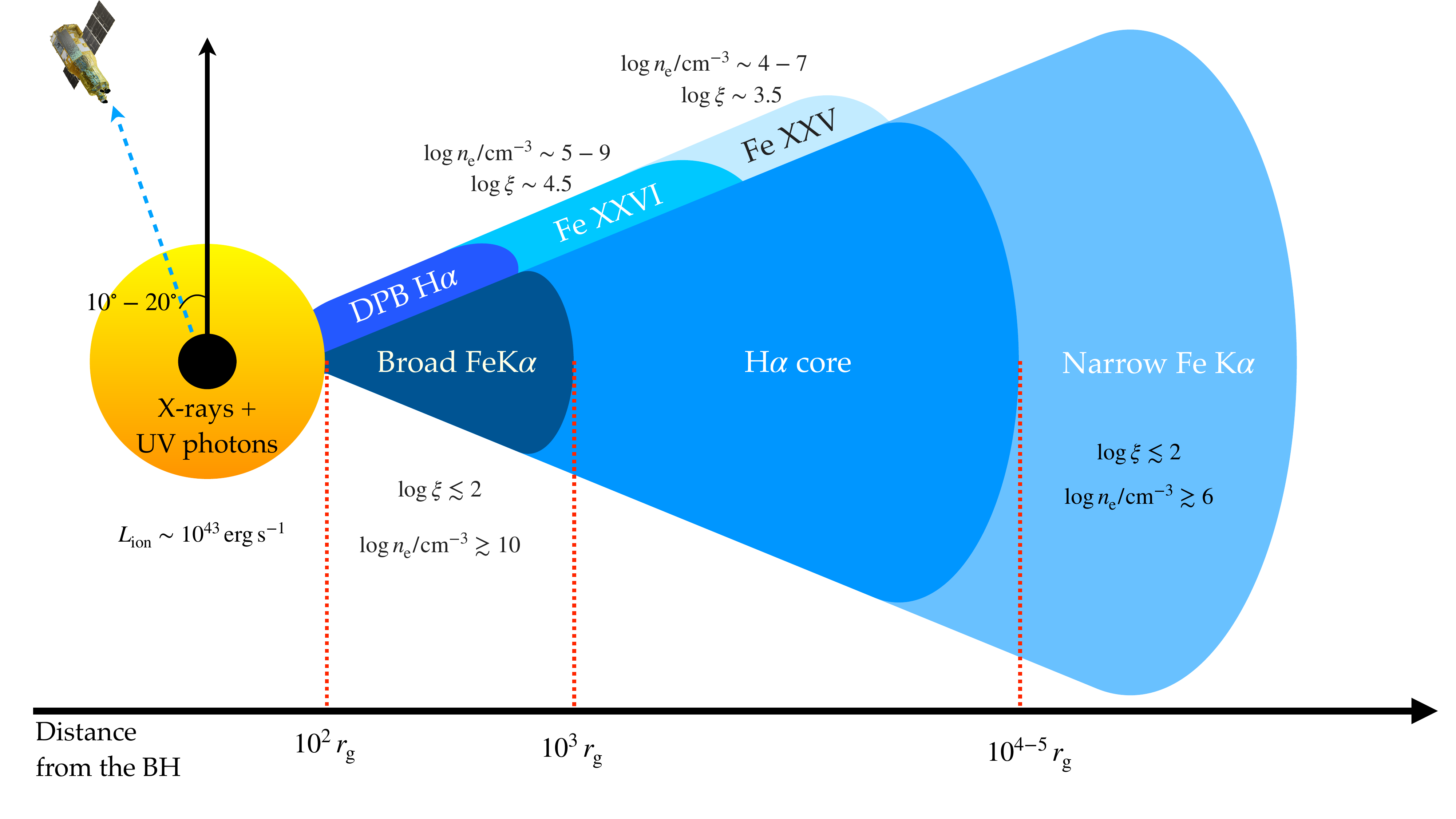}
  \caption{Schematic representation (not to scale), showing the radial stratification of the various emission lines reported in this work, under the assumption of photoionization by centrally concentrated X-ray and UV photons, and that the broadening of the lines is caused by orbital motion around the SMBH.}
  \label{fig:sketch}
\end{figure*}


Furthermore, two narrow features could be identified redwards of the \ion{Fe}{25} and \ion{Fe}{26} lines, at $\sim 6.56$\,keV and $\sim 6.82$\,keV, respectively, albeit with a low statistical significance. We modeled these lines by fixing their rest-frame energies at the rest-frame energies of the \ion{Fe}{25} resonant and \ion{Fe}{26}\,K$\alpha_1$ lines, and assumed the same redshift for both lines. We also fixed the width of the lines at 100\,\kms. The fit improved by only $\Delta C = -7$ for three additional free parameters. As discussed in Appendix\,\ref{appendix:line_sig}, these lines are significant with a 98\% confidence level ($\sim 2.3\sigma$). We find a redshift of $4600_{-110}^{+140}\,\kms$. If these lines are real, this may indicate the presence of a multi-phase inflow in the source.  Alternatively, if we assume that these lines are blueshifted low-ionization Fe\,K$\alpha$ and resonant \ion{Fe}{25} line, we find a velocity of $\sim -9000\,\kms$ for the line at $\sim 6.56$\,keV and $\sim -7000\,\kms$ for the line at $\sim 6.82$\,keV. The best-fit model after adding these two lines and the lines at $6.45$ and $6.75$\,keV, discussed previously, into the model is shown in Fig.\,\ref{fig:redshifted_lines}. We note that a hint of absorption at $\sim 6.52$\,keV appears in the data, though with low statistical significance. Deeper exposures may be required to confirm the presence of these features.

\subsection{A self-consistent picture}

We attempt in this section to present a complete and self-consistent picture inferred from the modeling the various features identified in the XRISM/Resolve data. We discuss two scenarios driven by the way the low-ionization Fe\,K$\alpha$ can be modeled.

In the previous sections, we estimated the distances of the emitting regions for each line, assuming that the Fe features detected in the XRISM/Resolve spectrum are produced by photoionization and that their observed broadening arises from the orbital motion of the gas around the black hole. In that scenario, we see a stratification of the line emitting regions as follows: the broad Fe\,K$\alpha$ and the DPB H$\alpha$ lines are produced at $\sim 100\,\rg$, the \ion{Fe}{26} lines are produced at $\sim 10^3\,\rg$, the \ion{Fe}{25} and the symmetric broad H$\alpha$ lines are produced at $\sim 10^4\,\rg$, and the narrow Fe\,K$\alpha$ line is produced at $\sim 10^5\,\rg$. Figure\,\ref{fig:sketch} shows a schematic representation of the radial distribution of the various lines, where the ionizing X-ray and UV photons are generated within $\sim 100\,\rg$ and the emission lines are distributed as discussed earlier. We note that this figure is not to scale and does not imply any assumptions on the geometry of the accretion flow below 100\,\rg\ and/or the geometry of the X-ray corona. In this figure the broad Fe\,K$\alpha$ and the DPB H$\alpha$ lines are produced at the same distance, in the inner part of the disk, which is connected to the region producing the broad symmetric core of H$\alpha$ line profile, at $R>10^3\,\rg$. The BLR emission is thought to be produced by gas with $\log (n_{\rm e}/{\rm cm^{-3}}) \gtrsim 10$, as estimated from the weakness of certain metastable and forbidden lines that become more prominent at lower gas density \citep[see e.g.,][and references therein]{Korista1997, Ulrich1997, Negrete2012, Schnorr-Muller2016, Abolmasov2017, Sniegowska2021}. The H$\alpha$ profile emerging from the outer (H$\alpha$ core) zone is single-peaked because of the large range in radii and a flat emissivity making up the core of the profile.

We use the definition of the ionization parameter to estimate approximately the electron density ($n_{\rm e}$) needed to produce each of the Fe lines at the corresponding distance ($R$) from the ionizing source. The ionization parameter is defined as $\xi = L_{\rm ion}/n_{\rm e} R^2$, where $L_{\rm ion}$ is the ionizing luminosity between $1-1000$\,Ry ($0.0136-13.6$\,keV). We assume an ionizing luminosity of $\sim 10^{43}\,\rm erg\,s^{-1}$, and that the Fe\,K$\alpha$, \ion{Fe}{25}, and \ion{Fe}{26} are produced at $\log (\xi/{\rm erg\,cm\,s^{-1}}) \lesssim 2, \sim 3.5,$ and $\sim 4.5$, respectively. 

The values of the ionization parameters correspond to the peak of the line emissivities estimated using \textsc{Cloudy} \citep{Cloudy25}. We obtain the following density estimates: $\log (n_{\rm e}/{\rm cm^{-3}}) \gtrsim 10$, $\sim 5-9$, $\sim 5-7$, $\sim 6$ for the different Fe lines according to their emitting radius. We note that these are approximate estimates given the uncertainties on the mass and the FWHMs, and they are not intended to give exact measurements. While most of these numbers are reasonable, the value of $\log (n_{\rm e}/{\rm cm^{-3}}) \sim 7$ is low compare to the density of the BLR emission. This suggests that the accretion flow itself is vertically stratified with an upper atmosphere with a low density, which could be also outflowing \citep[see e.g.,][]{Elitzur2014}. 

In this scenario, the ionization is stratified in the vertical direction as well as in the radial direction, with the densest part concentrating around the mid-plane and in the inner region. The gas could be then distributed in a way that gives rise in the inner regions to Fe\,k$\alpha$ and H$\alpha$ lines, although in different vertical layers, explaining the differences in the profiles of the two lines mentioned earlier. The outer region could then be emitting primarily  H$\alpha$ in the mid-plane, while the \ion{Fe}{25} and \ion{Fe}{26} lines are emitted by the upper layer of the disk atmosphere, explaining the need for lower density and higher ionization parameters. We note that the proposed pictures will not hold under the assumption of a collisionally ionized gas producing the \ion{Fe}{25} and \ion{Fe}{26} lines.

\begin{figure}
  \centering
  \includegraphics[width=\linewidth]{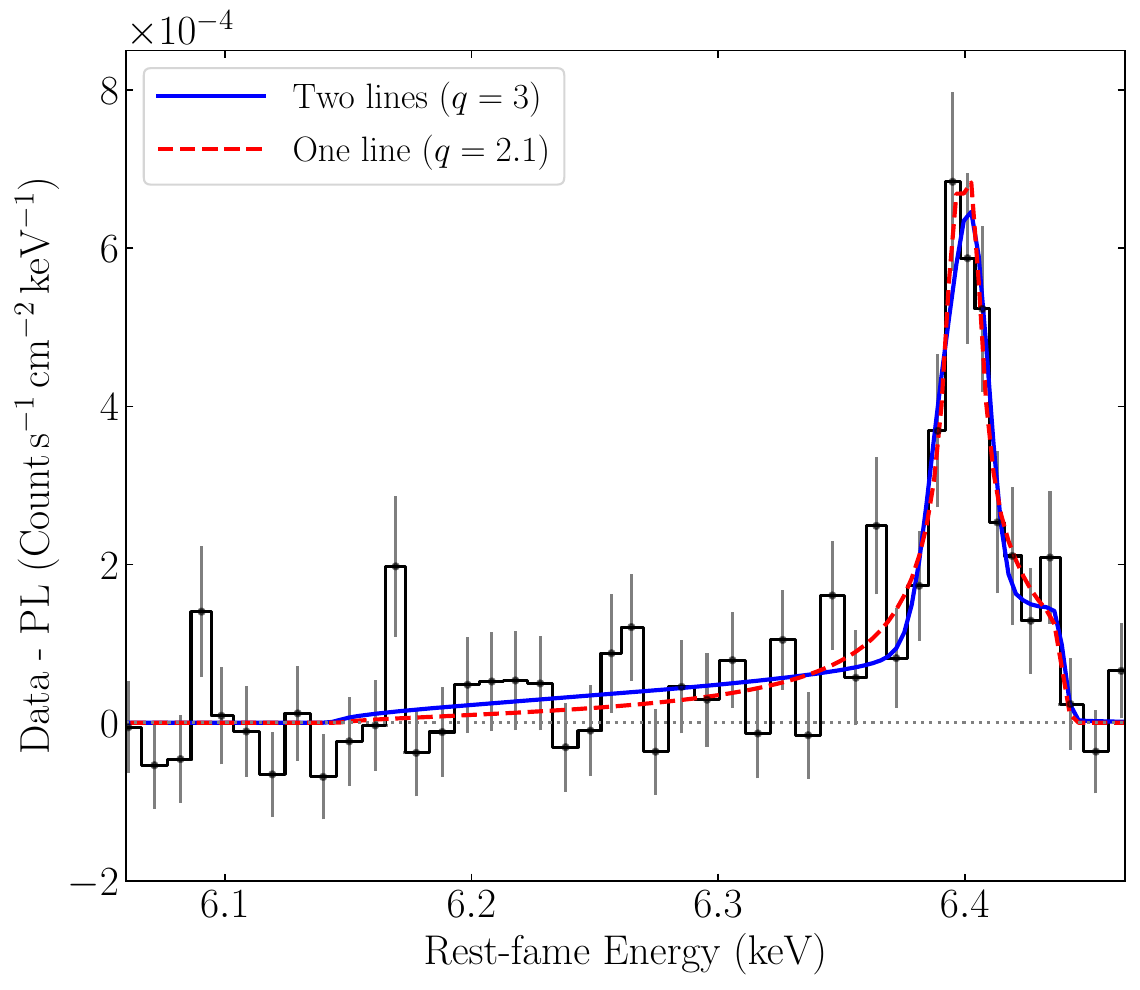}
  \caption{Best-fit of the low-ionization Fe\,K$\alpha$ assuming two lines (a broad base and a narrow core) with $q=3$ (solid blue line) and a single line with $q=2.1$ (red dashed line). We show the data after subtracting the power law.}
  \label{fig:two_solutions}
\end{figure}

In the previous sections we modeled separately the broad base and the narrow core of the low-ionization Fe\,K$\alpha$ at 6.4\,keV. However, these two features could also be modeled using a single \textsc{Diskline} feature extending from close to the black hole up to $\sim 10^5\,\rg$, with an emissivity index of $q\sim 2$ . A similar model has been found by \cite{Bogensberger2025} for Cen\,A. In our model, replacing $\textsc{GSmooth} \times \textsc{zFeKlor}_{\rm FeK\alpha} +  \textsc{diskline} $ in Eq.\,(\ref{diskmodel}) with only \textsc{Diskline}, leaving the emissivity profile free to vary results in an equally good fit ($C/{\rm dof}=510/524$). The best-fit emissivity profile of $q=2.1_{-0.2}^{+0.1}$, outer radius $\log (R_{\rm out}/\rg) = 5.2\pm 0.5$, and inclination $i = 11_{-2}^{+7}$ degree. We obtained a best-fit inner radius of 80\,\rg, with a $3\sigma$ upper limit of $R_{\rm in}<1000\,\rg$. We show in Fig.\,\ref{fig:two_solutions} the best-fits to the line assuming the sum of two lines with $q=3$ and a single line with $q=2.1$. This solution does not contradict the previous one consisting of two different lines. It presents a different perspective on the geometry of the system. The case of $q\sim 2$ is one of the rare emissivity profiles where the emission from both the inner part of the disk and the outer part can be detected in a single line profile. For lower emissivity indices, the line profile is dominated by the narrow core (i.e., the outer regions), while for higher values the inner radii (broad core) dominate the emission, unless two lines with independent normalization are used. This behavior has been known and discussed for a few decades now \citep[see e.g.,][]{Fabian89,Laor1991}. As pointed out by \cite{Laor1991}, a line with an emissivity profile of $q\sim 2$ is most sensitive to the outer edge. This is seen in our data by a well constrained $R_{\rm out}$ and only an upper limit on $R_{\rm in}$.

The two solutions strongly demonstrate that the contributions from both the inner and the outer radii to the line profile are needed to explain the observation. They both also result in consistent inner and outer radii, and inclination. The flatter than 3 emissivity profile can be an indication of flaring/concave accretion disk as shown by \cite{Blackman1999}, in their Fig.\,4. More recent general relativistic magneto-hydrodynamic (GRMHD) simulations also show that an emissivity profile consistent with $q\sim 2$ can be expected \citep[e.g.,][]{Kinch2016, Kinch2021}. These simulations also consider an accretion disk with a thickness increasing with radius, they also consider reprocessing in the upper layers of the disk which can also help flattening the emissivity profile. It could also be expected to have a broken power-law emissivity profile. However, the current data quality does not allow testing such a complicated model. Further observation of this source, as well as other AGN, would be of  great importance to study their line profiles and better understand the structure of the reprocessing material.

\subsection{\ngc\ as an intermediate case between LLAGN and Seyferts}

The XRISM/Resolve results place \ngc\ squarely between the torus-starved LINER M\,81$^{*}$ \citep[$\lambda_{\rm Edd} \simeq 10^{-5} $;][]{Ho1996}, where only a distant Fe\,K$\alpha$ core is detected \citep{Miller2025}, and the archetypal Seyfert NGC\,4151  \citep[$\lambda_{\rm Edd} \simeq 0.02 $;][]{Lub16}, which shows Fe\,K$\alpha$ emission from the torus down to $\sim 100\,R_{\rm g}$ \citep{Xrism2024ngc4151}. In NGC\,7213 a narrow core at $R\sim(0.7-3)\times10^{4}\,R_{\rm g}$ coexists with a disk-line at $R\sim40-300\,R_{\rm g}$, implying a radially stratified flow: an inner, low-luminosity annulus of a thin disk plus an outer region that resembles the torus/BLR complex of more luminous Seyferts. The narrow Fe\,K$\alpha$ radius in \ngc\ matches the dust-sublimation radius, yet its weak Compton hump and low Fe\,K$\beta$/Fe\,K$\alpha$ ratio echo the properties of M\,81$^{*}$. In addition, the EW of the narrow Fe\,K$\alpha$ in NGC\,7213 ($\rm EW = 32 \pm 6$\,eV) is similar to that found in M\,81$^{\ast}$ \citep[$\rm EW = 37_{-6}^{+7}$\,eV;][]{Miller2025}. We applied Eq.\,(\ref{eq:EW}) to M81$^{*}$, using the best-fit photon index $(\Gamma = 2)$ and column density ($N_{\rm H} = 1.6\times 10^{24}\,\rm cm^{-2}$) reported by \cite{Miller2025}. We find a torus covering fraction of 5\%. As for NGC\,4151, we estimate an EW of the distant Fe\,K$\alpha$ based on the best-fit parameters reported by \cite{Xrism2024ngc4151} to be $\sim 50$\,eV. For the best-fit photon index and column density \citep[$\Gamma = 1.7$ and $N_{\rm H} = 2\times 10^{23}\,\rm cm^{-2}$ as listed in Table\,1 of][]{Xrism2024ngc4151}, we find a torus covering fraction of 40\%. Thus, the torus covering fraction of \ngc\ lies between that of M81$^\ast$ and NGC\,4151. We therefore view \ngc\ as occupying an intermediate accretion regime in which the BLR and inner disk persist while the cold, optically thick torus may have already begun to dissipate. This further supports earlier predictions that the BLR and torus should disappear at low accretion rate \citep[see e.g.,][]{Nicastro2000, Laor2003, Elitzur2009}.

\section{Conclusions}
\label{sec:conclusion}

We have presented the first XRISM/Resolve observation of NGC\,7213, an intermediate/low accretion rate AGN ($\rm \lambda_{Edd} \sim 0.1–1\%$). The unprecedented spectral resolution of Resolve has enabled us to resolve the complex iron structure of this source, revealing radial stratification spanning nearly four orders of magnitude in distance from the central black hole.

Our key findings are as follows:

\begin{itemize}
  
    \item Multi-component Fe\,K$\alpha$ emission: The neutral Fe\,K$\alpha$ line consists of a narrow core ($\rm FWHM = 650\, km\, s^{-1}$) originating from distant, Compton-thin material, consistent with the dust sublimation radius. Superimposed on this is a broader, asymmetric component best described by relativistic disk emission from $R_{\rm in} \sim 100\,\rg$, viewed at a low inclination $i \sim 11\,\degr$. The broad Fe\,K$\alpha$ profile mirrors the double-peaked broad H$\alpha$ emission observed in our simultaneous SOAR spectrum. 

    \item Highly ionized iron emission: We detect resolved \ion{Fe}{25} (forbidden and resonant) and \ion{Fe}{26}\,K$\alpha_{1,2}$ emission lines with FWHM $\sim 2500\,\kms$ and $6000\,\kms$, respectively. Assuming Keplerian broadening, these lines originate at intermediate radii between the inner disk and optical BLR.

    \item Radial and vertical stratification: Under a photoionization scenario, the observed line properties suggest both radial stratification and vertical stratification, with high-ionization lines requiring lower densities ($\log (n_e/\rm cm^{-3}) \sim 5–7$) than expected for the BLR mid-plane ($\log (n_e/\rm cm^{-3}) > 10$), indicating emission from an extended disk atmosphere.

    \item Weak torus signature: The narrow Fe\,K$\alpha$ has a low equivalent width ($\rm EW = 32 \pm 6\, eV$), and the hard X-ray spectrum shows no significant Compton hump. These observations point to a low covering fraction ($f_{\rm cov} \sim 6–20\%$) and/or low column density (a few times $10^{23}\,\rm cm^{-2}$) torus, consistent with theoretical predictions for torus dissipation at low accretion rates.
\end{itemize}

Our results establish \ngc\ as occupying an intermediate accretion regime between classical Seyferts and  LLAGN. NGC\,7213 retains an inner accretion disk and functioning BLR, similar to higher-luminosity Seyferts. However, its torus shows clear signs of low covering fraction, similar to LLAGN. This supports theoretical predictions  for structural evolution with declining accretion rate and places the  critical $\lambda_{\rm Edd} \sim 0.01$ threshold where previous studies suggested the dusty torus begins to disappear.

The detection of both disk-like Fe\,K$\alpha$ emission and highly ionized \ion{Fe}{25} and \ion{Fe}{26} lines bridging the gap between the inner disk and BLR provides the first X-ray spectroscopic evidence, to our knowledge, for a continuous, radially stratified flow in an intermediate-accretion source. This suggests that the disk, disk wind, BLR, and torus may form a continuum rather than discrete regions.

Several aspects of our findings warrant further investigation. While our interpretation assumes photoionization drives the Fe XXV and Fe XXVI emission, collisional ionization in hot winds or shocks remains viable, and the \ion{Fe}{25} forbidden-to-resonant ratio ($0.9_{-0.5}^{+0.6}$) is not conclusive. We detect additional emission lines at 6.45\,keV and 6.75\,keV (each $\sim 3.3\sigma$ significant), plus weaker features at 6.56\,keV and 6.82\,keV ($\sim2 2.3\sigma$), which could indicate either inflow or outflow depending on the interpretations of the lines. The apparent broadening of \ion{Fe}{26} compared to Chandra/HETG upper limits suggests potential response to the factor-of-two luminosity increase, favoring photoionization, though confirmation requires multi-epoch monitoring. Future tests include: (1) coordinated optical-X-ray monitoring to reveal correlated Fe\,K$\alpha$/H$\alpha$ variability from disk structures and constrain vertical stratification through disparate timescales; (2) reverberation mapping to confirm photoionization and measure distances independently; (3) XRISM observations of additional sources with $\lambda_{\rm Edd} \lesssim 10^{-2}$ to establish whether NGC\,7213's apparent stratification and torus properties represent universal evolution or source-specific geometry; and (4) deeper XRISM exposures to confirm weak features, improve \ion{Fe}{25} diagnostics, and enable time-resolved spectroscopy.

Our results may have broad implications for AGN physics. The detection of highly ionized iron at intermediate radii from a low-density atmosphere supports thermally driven disk wind models where the BLR arises from the failed wind or disk atmosphere, with required density stratification consistent with such scenarios. The persistence of disk emission down to $\sim 100\,\rg$ at $\lambda_{\rm Edd} \sim 0.1–1\%$ constrains radiatively inefficient flow models. If a hot inner flow exists, its truncation radius must lie even closer in, or the disk remains thin throughout. The low covering fraction and weak Compton hump, intermediate between M\,81* and NGC\,4151, support gradual torus dissipation with decreasing accretion rate, with implications for AGN unification at low luminosities. In summary, XRISM/Resolve has revealed NGC\,7213 as a radially and vertically stratified system bridging LINERs and Seyferts. Future multi-wavelength campaigns targeting similar intermediate-accretion sources will be essential for understanding the transition between radiatively efficient and inefficient modes and mapping the innermost AGN structure across the full range of black hole feeding rates.


\textit{Facilities:} XRISM, XMM-Newton, NuSTAR, SOAR.

\section*{Acknowledgments}

We would like to thank the anonymous referee for their prompt and insightful feedback which helped improve the analysis. We would like to thank the SOC members of XRISM, NuSTAR, and XMM-Newton for coordinating the observations, as well as Anna Ogorzalek and the XRISM help-desk colleagues who assisted in the early stages of the data reduction. EK would like to thank Erin Kara and Dan Wilkins for fruitful discussions.

FN, EP, AL, FP, RS, MD, AC, GL, CP, EN, FT, GM, SB, CV and AM, acknowledge financial support from the Bando Ricerca Fondamentale INAF 2023, Large Program 1.05.23.01.06 (``The XRISM-to-XIFU (X2X) Agreement and Beyond: entering a new Era of High Resolution X-Ray Spectroscopy''). FT, FN, ML, MD and MC, acknowledge financial support from PRIN MUR 2022 DRAGON 2022K9N5B4. EB acknowledges financial support from INAF under the Large Grant 2022 ``The metal circle: a new sharp view of the baryon cycle up to Cosmic Dawn with the latest generation IFU facilities'', the GO grant ``A JWST/MIRI MIRACLE: Mid-IR Activity of Circumnuclear Line Emission'' and the ``Ricerca Fondamentale 2024'' program (mini-grant 1.05.24.07.01). VEG acknowledges funding under NASA contract 80NSSC24K1403.  FP acknowledges financial support from the Bando Ricerca Fondamentale INAF 2023. CP acknowledges funding from PRIN MUR 2022 SEAWIND 2022Y2T94C, supported by European Union - Next Generation EU, Mission 4 Component 1 CUP C53D23001330006. GP acknowledges support from the European Research Council (ERC) under the European Union's Horizon 2020 research and innovation program HotMilk (grant agreement No. 865637), support from Bando per il Finanziamento della Ricerca Fondamentale 2022 del'Istituto Nazionale di Astrofisica (INAF): GO Large program and from the Framework per l'Attrazione e il Rafforzamento delle Eccellenze (FARE) per la ricerca in Italia (R20L5S39T9). RS acknowledges funding from the CAS-ANID grant number CAS220016.

This research has made use of data and/or software provided by the High Energy Astrophysics Science Archive Research Center (HEASARC), which is a service of the Astrophysics Science Division at NASA/GSFC and the High Energy Astrophysics Division of the Smithsonian Astrophysical Observatory. This research made use of pandas \citep{McKinney_2010, McKinney_2011} and Astropy, a community-developed core Python package for Astronomy \citep{2018AJ....156..123A, 2013A&A...558A..33A}. This research made use of XSPEC \citep{1996ASPC..101...17A}. This research made use of SciPy \citep{Virtanen_2020}, matplotlib, a Python library for publication quality graphics \citep{Hunter:2007} and NumPy \citep{harris2020array}. This research is based on observations obtained with XRISM, a JAXA/NASA collaborative mission, with ESA participation, XMM-Newton, an ESA science mission with instruments and contributions directly funded by ESA Member States and NASA. NuSTAR mission, a project led by the California Institute of Technology, managed by the Jet Propulsion Laboratory, and funded by the National Aeronautics and Space Administration. Data analysis was performed using the NuSTAR Data Analysis Software (NuSTARDAS), jointly developed by the ASI Science Data Center (SSDC, Italy) and the California Institute of Technology (USA).

\appendix 

\section{The effect of the cutoff rigidity}
\label{appendix:COR}

\begin{figure*}
    \centering
    \includegraphics[width=0.5\linewidth]{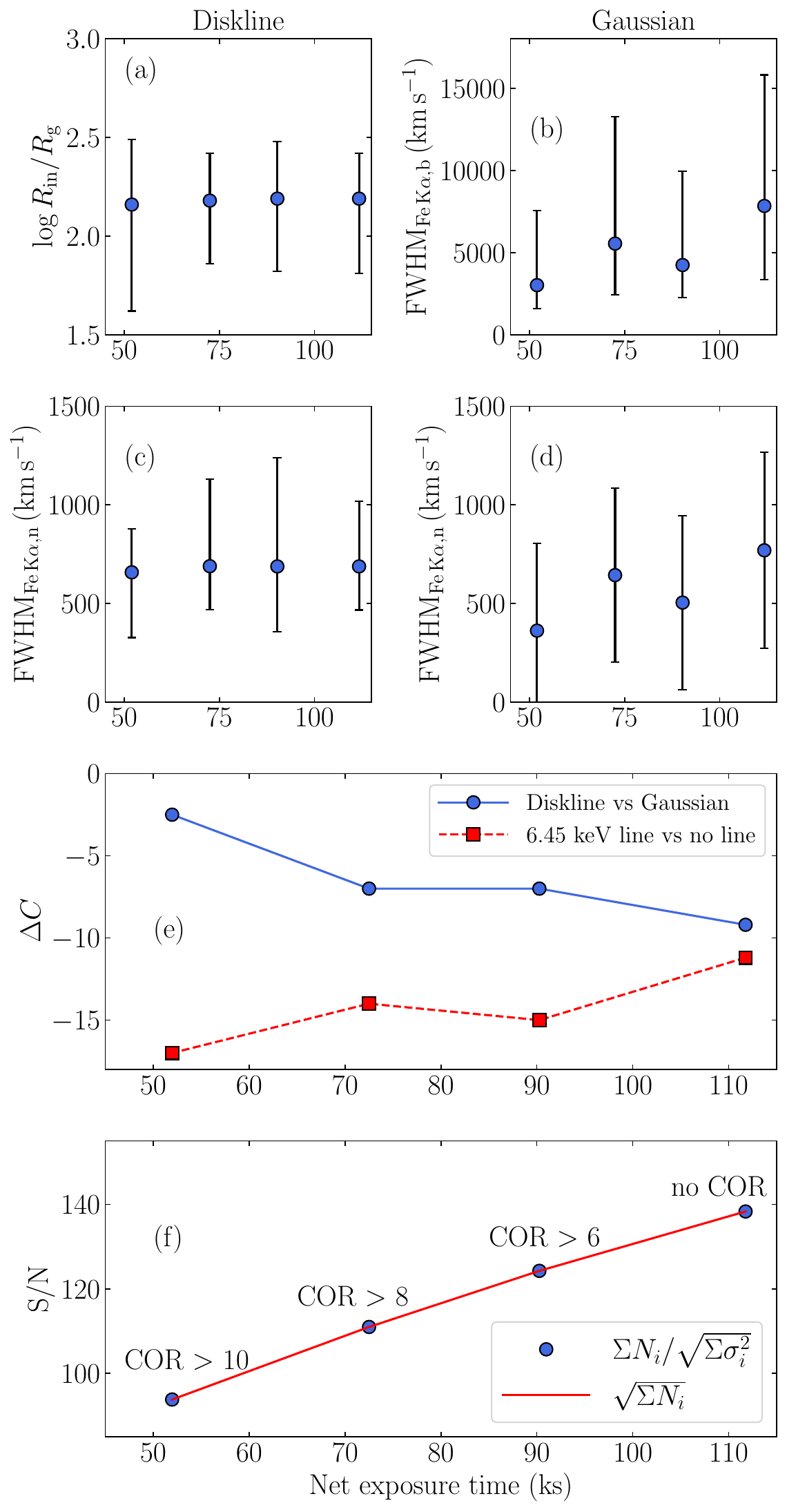}
    \caption{Effect of the COR on the best-fit parameters. Left and right panels correspond to the best-fit parameters obtained for modeling the broad Fe\,K$\alpha$ with a \textsc{diskline} and symmetric Gaussian-broadened line, respectively. Panels (a) and (c) show the inner radius for the \textsc{diskline} and the FWHM of the narrow Fe\,K$\alpha$, respectively. Panels (b) and (d) show the FWHM of the broad and narrow Fe\,K$\alpha$ lines, respectively. Panel (e) shows the improvement in fits statistics for a \textsc{diskline} with respect to a Gaussian-broadened line (blue circles) and the improvement obtained by including the emission line at 6.45\,keV (red squares) for different COR values. Panel (f) shows the signal to noise ratio of the full spectrum in the $5.5-7.5$\,keV range (see text for details).}
    \label{fig:CORfits}
\end{figure*}

We explore in this section the effect of the cutoff rigidity on our results. We extracted XRISM/Resolve spectra assuming no COR, $\rm COR > 6$, $8$, and $10$. The net exposure time decreases for $ 111$\,ks for no COR to 52\,ks for $\rm COR > 10$. We fitted each of the spectra with the model presented in Eq. (\ref{gaussmodel}) which models that the broad low-ionization Fe\,K$\alpha$ line with a Gaussian-broadened line and with a \textsc{Diskline}. We show in Fig.\,\ref{fig:CORfits} the best-fit parameters as a function of the net exposure time. For the \textsc{Diskline} model we show the inner radius and FWHM of the narrow Fe\,K$\alpha$ (panels a and c, respectively). For the Gaussian-broadened model we show the FWHMs of the broad and narrow   Fe\,K$\alpha$ (panels b and d, respectively). This figure shows that the properties of the Fe\,K$\alpha$ lines are consistent for the diffrent COR choices. In Panel (e), we show the improvement of the fit ($\Delta C$) obtained by using \textsc{diskline} instead of a symmetric broadening (blue circles). The fit improvement due to the \textsc{diskline} increases for larger exposures. We note that this is not a measure to whether the broad line exists or not. It is to test the ability to identify an asymmetric versus a symmetric line profile. We also show the improvement of the fit when a Gaussian line at 6.45\,keV is included (red squares). This has an opposite trend with respect to the broad line. We believe that this opposite behavior is caused by the effect of the COR. For the broad line, the distinction between an asymmetric and a symmetric line requires a higher number of counts (thus a larger overall S/N). We show the overall S/N in the $5.5-7.5$\,keV range in Panel (f). We estimate S/N as follows:
\begin{equation}
     S/N = \frac{\sum_{i} N_i}{\sqrt{\sum_i \sigma_i^2}},
\end{equation}
\noindent where $N_i$ and $\sigma_i$ are the number of counts and their corresponding uncertainty in each channel. As Panel (f) shows, this is equivalent for $S/N = \sqrt{\sum_i N_i}$, which is expected for Poisson statistics. However, in the case of the narrow line, excluding low COR intervals enables a better detection of the weak features. In any case, even for the case of no COR filtering, the improvement of the fit caused by the addition of the narrow line is still significantly high ($\Delta C < - 10$). This exercise justifies our choice of $\rm COR > 6$, as an intermediate case where the loss in exposure is only $\sim 10\%$.

\section{MCMC convergence}
\label{appendix:MCMC}

Standard convergence diagnostics (Geweke $z$-scores $|z| < 0.3$, stable cumulative means and variances) together with an effective sample size of $\sim 7\times 10^4$ confirm that the retained samples provide an adequate representation of the target posterior. For all parameter–space visualizations we plot the raw post-burn chain, whose point density is already proportional to the posterior. Quantities derived from the posterior (such as synthetic line profiles or marginal percentiles) are computed from a thinned subset limited to samples with $\Delta C\le10$. Each retained sample has an associated $C$-statistic stored by \textsc{XSPEC}; we restore the exact posterior measure by assigning the weight $w_k = \exp[-0.5(C_k - C_{\rm min}) ]$ and re-normalizing so that $\sum_k w_k=1$. This weighting reproduces the full posterior integrals at negligible computational cost. Weighted percentiles of the model flux in every energy bin yield the median spectrum and its 68\% and 95\% credibility bands (right-hand panel of Fig.\,\ref{fig:XRISM_spectra_fit}). The procedure automatically propagates the full parameter covariance without additional error analysis.  Marginal medians and 90\% credibility limits are reported in Table\,\ref{tab:bestfit-xrism} and defined in the same way.

\section{Line significance}
\label{appendix:line_sig}

As it is already known, it is not straightforward to assess the significance of a spectral component using $C$-stat. For that reason we ran a set of simulations using the same response files and exposure time (90\,ks) as the observation studied in this work.

\subsection{\textsc{diskline} vs symmetric line}

As discussed in Section\,\ref{section:Feregion}, the best-fit improves by $\Delta C = C_{\rm diskline} -- C_{\rm symmetric} = -7$ using an asymmetric \textsc{diskline} compared to a symmetric Gaussian-broadened emission line. We used the best-fit model in which the broad low-ionization Fe\,K$\alpha$ is modeled with $\textsc{Gsmooth}\times\textsc{zFeKlor}$. Using the \textsc{xspec} command \textsc{fakeit}, we produced 1000 simulated spectra which we fitted using the same model, and also using \textsc{diskline}. In the latter case, we restricted the parameter space to $R_{\rm in} < 500\,\rg$ and $i < 20\degr$, to be consistent with the best-fit results, favoring also an asymmetric profile. Allowing $R_{\rm in}$ to reach larger values would result in more symmetric line profiles. We found that when simulating with a symmetric line an asymmetric model could provide a better fit with $\Delta C \leq -7$ in only 9 cases out of 1000. This is equivalent to confidence level of 99.1\% (equivalent to 2.6$\sigma$) that the asymmetric is not produced by chance due to the noise in the data. We also repeated the same exercise by simulation an asymmetric line while fitting with both an asymmetric and a symmetric model. In this case, we find that the median of the $\Delta C$ is $-7.6$, which is consistent with the value of $-7$ that we found from the observations. In addition, we found that when simulating a asymmetric line, a symmetric line could be preferred ($\Delta C >0$) in 5\% of the cases.

\subsection{Gaussian lines at 6.45\,keV and 6.75\,keV}

To estimate the significance of the emission lines at 6.45\,keV and 6.75\,keV, we also ran a set of 1000 simulations assuming the best-fit model presented in Section\,\ref{section:Feregion}, where we set the normalizations of these two lines to be zero. Then, we fitted the simulated spectra using a model with normalizations fixed at 0, and then by letting the normalizations free to vary. In this case, we can assess the probability of finding those lines due to noise in the data. We find that, for both lines, only 1/1000 case where $\Delta C = C_{\rm line} - C_{\rm no\, line}$ is smaller than the values we found from the observations (see Table\,\ref{tab:bestfit-xrism}). This indicates that the presence of these lines is significant at $99.9\%$ ($\sim 3.3\sigma$).

\subsection{Weak Gaussian lines at 6.56\,keV and 6.82\,keV}
We repeated the same exercise to assess the significance of the lines discussed in Section\,\ref{sec:gaussians} at $\sim 6.56$ and 6.82\,keV (found with $\Delta C =  C_{\rm line} - C_{\rm no\, line} = -7$). Similarly to the previous section, we ran simulations assuming the normalizations are zero. We fitted with normalizations fixed at zero, and free to vary. In this case, we find that these lines could be found in the simulated spectra with $\Delta C < -7$ in 20 simulations out of 1000. This indicates a confidence level of 98\% ($\sim 2.3\sigma$).

\bibliography{references}{}
\bibliographystyle{aasjournalv7}

\end{document}